\documentclass{jfm}
\usepackage{esvect}
\usepackage{graphicx}
\usepackage{epstopdf, epsfig}
\usepackage{float}
\usepackage{amsfonts}
\usepackage{xcolor}

\usepackage{mathtools}
\usepackage{amsmath}
\usepackage{multicol}
\usepackage{amsmath}

\allowdisplaybreaks 

\shorttitle{Laminar/turbulent interface and energy transfer between scales}
\shortauthor{H. Yao and G. Papadakis}

\title{On the role of laminar/turbulent interface on energy transfer between scales in bypass transition}

\author{H. Yao\aff{1}
 \and G. Papadakis\aff{1} \corresp{\email{g.papadakis@imperial.ac.uk}}}

\affiliation{\aff{1}Department of Aeronautics, Imperial College London, London SW7 2AZ, UK}

\begin{document}

\maketitle

\begin{abstract}
We investigate the role of laminar/turbulent interface in the interscale energy transfer in a boundary layer undergoing bypass transition, with the aid of the Karman-Howarth-Monin-Hill (KHMH) equation.  A local binary  indicator function is used to detect the interface and employed subsequently to define two-point intermittencies.  These are used to decompose the standard-averaged interscale and interspace energy fluxes into conditionally-averaged  components. We find that the inverse cascade in the streamwise direction reported in an earlier work arises due to events across the downstream or upstream interfaces (head or tail respectively) of a turbulent spot. However, the three-dimensional energy flux maps reveal significant differences between these two regions: in the downstream interface, inverse cascade is stronger and dominant over a larger range of streamwise and spanwise separations. We explain this finding by considering a propagating spot of simplified shape as it crosses a fixed streamwise location. We derive also the conditionally-averaged KHMH equation, thus generalising similar equations for single-point statistics to two-point statistics. We compare the three-dimensional maps of the conditionally-averaged production and total energy flux within turbulent spots against the maps of standard-averaged quantities within the fully turbulent region. The results indicate remarkable dynamical similarities between turbulent spots and the fully turbulent region for two-point statistics. This has been known only for single-point quantities, and we show here that the similarity extends to two-point quantities as well. 
\end{abstract}

\begin{keywords}
\end{keywords}

\section{Introduction}
\subsection{Bypass transition}
Transition to turbulence that does not involve linear instability paths, such as Tollmien–Schlichting waves, is called bypass transition \citep{morkovin1969many}. This type of transition can be triggered by high levels of free-stream turbulence, surface roughness, etc. In the case of free-stream turbulence, which is the triggering mechanism considered in this paper, bypass transition comprises three stages. In the first stage, low-frequency fluctuations from the free-stream penetrate inside the boundary layer, forming high and low-speed streaks, while high-frequency fluctuations remain in the free-stream due to shear sheltering \citep{leib1999effect, hunt1999perturbed, zaki2009shear}. In the second stage, the streaks breakdown to intermittent turbulent patches (or spots)  due to secondary instability \citep{andersson2001breakdown,vaughan2011stability}, while in the final stage the spots propagate and merge, forming a fully turbulent region. More details can be found in the review papers of \cite{durbin2007transition,zaki2013streaks,durbin2017perspectives}.

Most previous investigations of the structural details of turbulent spots, such as shape, propagation speed, growth rate etc.\ have employed analysis of single point statistics, see \cite{emmons1951laminar,wygnanski1976turbulent,Cantwell_coles_dimotakis_1978,perry1981visual,singer1996characteristics, Nolan_Zaki_2013} and more recently \cite{wang2021early,Wang_Choi_Gaster_Atkin_Borodulin_Kachanov_2022}. This type of analysis however cannot capture the underlying physical mechanisms that explain the spot growth and the amalgamation process of smaller spots to form larger turbulent patches as transition progresses. In order to study this process in more detail, an analysis of two-point statistics is required.  The second order structure function at point $X_i$ is defined as the second moment of the fluctuating velocity difference at points $x^\pm_i=X_i\pm \frac{1}{2}r_i$, i.e. $\overline{dq^2}(X_i,r_i)=\overline{(u'^+_i-u'^-_i)^2}$, where the overbar denotes time-averaging. The volume integral of $\overline{dq^2}$ over a sphere of radius $r=|r_i|$ (divided by the volume of the sphere) represents physically the energy of eddies located at $X_i$ that have size (or scale) less than $r$; this is also known as scale energy, for details see \cite{davidson2015turbulence}. This is the appropriate quantity to study in order to better understand the process of spots growth and merging. 

The transport equation of $\overline{{dq^2}}(X_i,r_i)$ is known as the Karman-Howarth-Monin-Hill (KHMH) equation. It contains all the physical mechanisms that determine the energy contained within eddies of scale less than $r$, such as transfer of energy in scale space (i.e.\ from smaller or larger scales), transfer of energy in physical space, production (due to mean shear), dissipation (due to viscosity), etc. It was first derived by \cite{Karman_Howarth_1938_new} for homogeneous isotropic turbulence (HIT) using the two-point velocity correlation tensor, $\overline{u'^+_iu'^-_j}$,  and later reformulated in terms of structure function  $\overline{dq^2}(r_i)$ by \cite{Kolmogorov_1941b_NEW}. The equation was used to prove the famous -4/5'th law, that links the interscale energy flux, the separation between the two points, and the dissipation rate. For homogeneous, isotropic turbulence the interscale flux is always negative, i.e.\ energy is transferred from large to small scales; this is known as forward cascade. The most general form of the KHMH equation, applicable to inhomogeneous and anisotropic flows, was derived directly from Navier-Stokes equations by \cite{hill2002exact}. The equation was applied recently to transitional boundary layers and demonstrated strong inverse cascade in the transition region, especially in the streamwise direction, see \cite{yao2022analysis}. Analysis of instantaneous velocity fields and flux vectors revealed that the inverse cascade was related to the growth of turbulent spots. 

The KHMH equation has been applied to several other flow settings, such as homogeneous shear flow \citep{Casciola_Gualtieri_Benzi_Piva_2003}, channel flow  \citep{marati2004energy, cimarelli2013paths,cimarelli2015sources,cimarelli2016cascades}, Von Karman flow \citep{dubrulle2019beyond,knutsen2020inter}, temporal planar jet \citep{cimarelli2021spatially}, wake behind a square cylinder \citep{portela_papadakis_vassilicos_2017, portela_papadakis_vassilicos_2020} and has revealed convoluted paths of interscale energy fluxes with a mixture of forward and inverse cascade. In all these investigations, the points $x^\pm_i$ used to define $\overline{dq^2}$ are immersed within a turbulent flow. This is not the case however in transitional flows, where for a fixed streamwise location $X_i$ and separation $r_i$, the two points will experience different flow conditions as a spot propagates, for example  $x^\pm_i$ may be within the laminar region, straddle the laminar/turbulent interface, or be within a turbulent patch. It is therefore difficult to explain the origin of the inverse cascade found in \cite{yao2022analysis}. For example, does it arise from the turbulent conditions within the spot? What is the role of the laminar/turbulent interface?  Is there competition between the different flow conditions, i.e.\ do some lead to forward and others to inverse cascade?        

To answer these questions in the present paper we perform conditional averaging of the interscale energy fluxes based on the state of the two points i.e.\ whether they experience laminar or turbulent flow conditions. This process clearly elucidates the effect of different flow states and the role of the laminar/turbulent interface.  As will be seen latter, it even characterises the separate roles of the downstream and upstream interfaces (head or tail respectively) of the spot. We also examine the production term of the KHMH equation conditioned on turbulent events within a turbulent patch, and compare it with the production term when the two points are located within the fully turbulent region. Similar comparisons have been made for single-point quantities, such as turbulent kinetic energy, see \cite{marxen2019turbulence}. We derive also the conditionally-averaged form of the KHMH equation, which is analogous to the conditionally-averaged turbulent kinetic energy equation. 

The work of \cite{zhou2020energy} has some similarities, but also significant differences, with the present work. The authors studied the energy cascade across the turbulent/non-turbulent (TNTI) interface at one axial position of an axisymmetric turbulent wake. They found that the interscale energy transfer at the vicinity of the interface is from small to large scales (inverse cascade) in directions close to the interface’s tangent plane where motions are predominantly stretching, but from large to small scales (forward cascade) in the other directions where motions are predominantly compressive. This reflects the fundamental mechanism that sustains the TNTI, i.e.\ fluid is entrained from the irrotational region and the wake grows due to turbulent diffusion, see schematic 3(d) in  \cite{zhou2020energy}. This mechanism however is different compared to the one that determines the growth of spots in a  transitional boundary layer. The authors also did not perform conditional analysis, because the midpoint $X_i$ was located at the interface (thus was not fixed in the cross-stream direction), and the two points  $x^\pm_i$ always straddled the TNTI. 

The paper is organised as follows; in section  \S\ref{sec:test_cases} the bypass transition case is briefly presented, in \S\ref{sec:standard_KHMH} we summarise the derivation of the standard time-averaged form of the KHMH equation, while in \S\ref{sec:conditionally_KHMH_equation} we derive the conditionally-averaged form; this is followed by the conditional decomposition of the energy fluxes in \S\ref{sec:cond_decompostition}. The next two sections present the results; in  \S\ref{sec:evolution_cond_averaged_fluxes} maps of the conditionally-averaged non-linear energy fluxes are shown  (focusing on the flux across the laminar/turbulent interface), while in \S\ref{sec:conditional_averaged_production} we compare the two-point energy production and total flux (both conditioned on turbulent events) to the corresponding quantities in the fully turbulent region. We conclude in \S\ref{sec:conclusions}. 

\section{Details of the test case examined}\label{sec:test_cases}
We consider the transition of a boundary layer developing on a flat plate due to free-stream turbulence. At the inlet of the computational domain, a random velocity field is superimposed on the Blasius velocity profile. In the free-stream,  the random field follows a von-Karman spectrum with turbulence intensity $3.4\%$ and integral length scale $L_{11}=5L_0$, where $L_0 = \sqrt{\nu X_0/U_\infty}$ is the Blasius similarity variable, $X_0$ is the distance of the inlet of the domain from the leading edge of the plate, $\nu$ is the kinematic viscosity and $U_\infty$ is the free-stream velocity. The inlet Reynolds number is $Re_{L_0}= 160$ (or $Re_\theta = 110$ based on momentum thickness). 

The size of the computational domain is $(3000 \times 200 \times 150)L_0$, with the number of cells $2049\times192\times169$ in the streamwise ($X$), wall-normal ($Y$), and spanwise ($Z$) directions.  Velocities are denoted as $u$, $v$, $w$ in the $X,Y,Z$ directions respectively. This notation is used interchangeably with the indexed notation $X_i$ and $u_i$ (with $i=1,2,3$), for example $X_2=Y$ and $u_2=v$. The spacing is uniform in the streamwise and spanwise directions, with $\Delta x^{+}_{max}\approx 11.78$ and $\Delta z^{+}_{max}\approx 7.14$, where ${max}$ represents the maximum value (located in the fully turbulent region). In the wall-normal direction, grid spacing increases gradually; $y^+$ at the centroid of the first cell close to the wall is around 0.24. The generated DNS database contains 350 uncorrelated snapshots. The results have been validated against the T3A experimental data \citep{roach1990influence}. More details about the computational method and comparison of velocity profiles (mean and RMS) against experiments can be found in \cite{Yao_Portela_Papadakis_2020}. 

\begin{figure}
    \centering
  \includegraphics[scale=0.5]{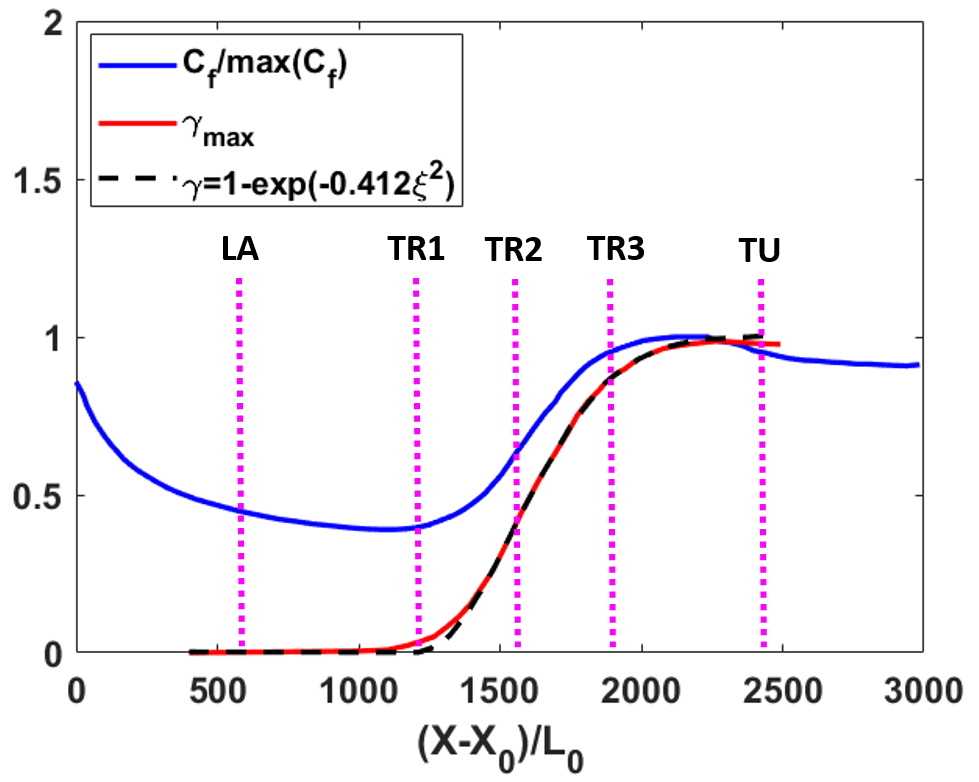}
    \caption{Normalised skin friction coefficient, $C_f/\max(C_f)$ (blue solid line), maximum intermittency $\gamma_{max}$ (red solid line), and $\gamma(\xi)$ from the formula of  \cite{narasimha1985laminar}, where  $\xi=(X-X_s)/(X_{\gamma=0.75}-X_{\gamma=0.25})$ and $X_s-X_0=1100L_0$ is the location where transition starts (black dash line) against streamwise distance $(X-X_0)/L_0$. The purple vertical lines are located in the laminar  ($LA=540L_0+X_0$), transitional ($TR1,TR2,TR3=1215,1515,1815L_0+X_0$) and fully turbulent  ($TU=2415L_0+X_0$) regions. }
    \label{Cf_sch}
\end{figure}

For future reference, the normalised skin friction coefficient, $C_f/\max(C_f)$, and the maximum value of the time- and spanwise-averaged intermittency are plotted in figure \ref{Cf_sch}. For the methodology used to compute the intermittency refer to the aforementioned paper and also to section \ref{sec:conditional_operations}. Vertical lines indicate the streamwise locations in the laminar ($LA$), transitional ($TR1,TR2, TR3$) and fully turbulent regions ($TU$) where velocity data are extracted in order to compute the interscale fluxes. 

\section{Standard time-averaged KHMH equation}\label{sec:standard_KHMH}
In this section, the form of the standard time-averaged KHMH equation is presented. The basic steps of the derivation are sketched below; more details can be found in \cite{hill2002exact}. Similarities and differences with the conditionally-averaged form are presented and discussed in section \ref{sec:conditionally_KHMH_equation}. 

\begin{figure}
    \centering
  \includegraphics[scale=0.5]{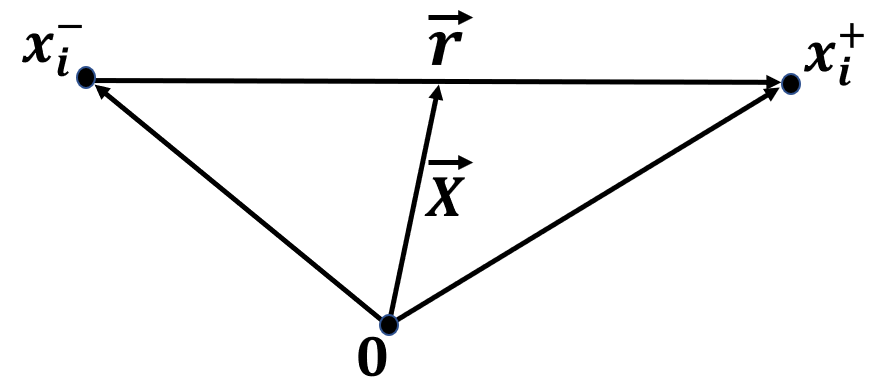}
    \caption{Sketch that shows the position vector  $\protect\vv{X}=(X_1,X_2,X_3)$ of the mid point, and the separation vector $\protect\vv{r}=(r_1,r_2,r_3)$ between the two points $\protect\vv{x^+}=(x^+_1,x^+_2,x^+_3)$ and $\protect\vv{x^-}=(x^-_1,x^-_2,x^-_3)$.}
   \label{fig:X_r_sch}
\end{figure}

We start with the Navier-Stokes equations at two points $x_i^+$ and $x_i^-$ (see sketch \ref{fig:X_r_sch}),
\begin{subequations}
 \begin{align}
 \frac{\partial u^+_i}{\partial t} 
 + u^+_j \frac{\partial u^+_i}{\partial x^+_j}
 &=-\frac{\partial p^+}{\partial x^+_i} + \nu \frac{\partial^2 u^+_i}{\partial x_j^+ \partial x_j^+}, \label{eq:momentum_at_x+}
 \\
  \frac{\partial u^-_i}{\partial t} 
 + u^-_j \frac{\partial u^-_i}{\partial x^-_j}
 &=-\frac{\partial p^-}{\partial x^-_i} + \nu \frac{\partial^2 u^-_i}{\partial x_j^- \partial x_j^-}
 \label{eq:momentum_at_x-}
\end{align}
\label{eq:momentum_at_x+_and_x-}
\end{subequations}
and define the time- and spanwise-averaged velocities as usual, 
\begin{subequations}
\begin{align}
&U^+_i(X,Y)=\overline{u^+_i}= \frac{1}{\Delta T L_z} \int_0^{\Delta T} \int_0^{L_z} u_i^+ \: dz dt, \\
&U^-_i(X,Y)=\overline{u^-_i}= \frac{1}{\Delta T L_z} \int_0^{\Delta T} \int_0^{L_z} u_i^- \: dz dt.
\end{align}
\label{eq:standard_averaging}
\end{subequations}
In the following, we use an overbar $\overline{(\quad)}$ to denote the standard averaging operation in the time and Z-direction, as defined by \eqref{eq:standard_averaging}. Velocity fluctuations around $U^+_i$ and $U^+_i$ are denoted using primes $(\quad)'$, i.e.\
\begin{equation}
u'^+_i=u^+_i-U^+_i,  \quad u'^-_i=u^-_i-U^-_i,
\label{eq:BB2_standard}
\end{equation}
and fluctuating velocity differences are defined as,
\begin{equation}
du_i' \equiv du_i-dU_i,
\label{eq:BB1_standard}
\end{equation}
\noindent where $du_i=u_i^+-u_i^-$ and $dU_i=\overline{du_i}=U^+_i-U^+_i$. It is straightforward to prove that $\overline{du_i'}=0$. 

Subtracting equation \eqref{eq:momentum_at_x-} from \eqref{eq:momentum_at_x+}, multiplying each term by $2du_i'$, and applying the time- and spanwise- averaging operation defined in \eqref{eq:standard_averaging} we obtain,
 \begin{equation}
 \begin{aligned}
&  \underbrace{
  \overline{{2du_i'}\frac{\partial du_i}{\partial t}}}_{\text{Transient term}}
 + 
  \underbrace{
  \overline{2du_i'u^+_j \frac{\partial du_i}{\partial x^+_j} 
 +
 2du_i'u^-_j \frac{\partial du_i}{\partial x^-_j}}}_{\text{Non-linear term}}
 = \\
&  \underbrace{
   \overline{{-2du_i'}\left(\frac{\partial dp}{\partial x^+_i} 
 -
 \frac{\partial dp}{\partial x^-_i}\right)}} _{\text{Pressure-velocity correlation}}
 +
 \underbrace{
  \overline{{2 \nu du_i'}\frac{\partial^2 du_i}{\partial x_j^+ \partial x_j^+} 
+
2\nu du_i'\frac{\partial^2 du_i}{\partial x_j^- \partial x_j^-}}
}_{\text{Viscous term}},
\label{eq:standard_time_average_Fin2}
\end{aligned}
\end{equation}
where we have used $\dfrac{\partial u^+_i}{\partial x^-_j}=0$ and $\dfrac{\partial u^-_i}{\partial x^+_j}=0$ (because $x^+_j$ and $x^-_j$ are independent variables).  

We now define the second order structure function as $\overline{dq^2} =\overline{(du'_i)^2}=\overline{(u'^+_i-u'^-_i)^2}$. This function has six dimensions, three in physical space ($X_i$) and three in scale space ($r_i$). In the particular case examine in this paper, due to the homogeneity in the spanwise direction, there are only two dimensions in physical space. The integral of $\overline{dq^2}$ in a sphere of radius $r=|r_i|$ (divided with the volume of the sphere) represents the energy of eddies with size smaller than $r=|r_i|$, see \cite{davidson2015turbulence}; thus $\overline{dq^2}$ is usually referred to as scale energy. 

We seek the transport equation of $\overline{dq^2}$ in the physical and scale spaces. Applying the variable transformation $X_i=0.5\left(x^+_i + x^-_i\right)$ and  $r_i = x^+_i - x^-_i$ and the definitions \eqref{eq:BB2_standard} and \eqref{eq:BB1_standard} into \eqref{eq:standard_time_average_Fin2}, after some algebra we get the following standard KHMH equation for $\overline{dq^2}$,
 \begin{equation}
 \begin{aligned}
& \underbrace{\overline{\frac{\partial  dq^2}{\partial t}
 }}_{\text{Transient term}} 
 + 
 \underbrace{\overline{ U^*_{j}\frac{\partial dq^2}{\partial X_j} }}_{\text{Mean flow advection}}
 +
  \underbrace{\overline{ u'^*_{j}\frac{\partial dq^2}{\partial X_j} }}_{\text{Turbulent advection}}
+
 \underbrace{\overline{ dU_j \frac{\partial dq^2}{\partial r_j} }}_{\text{Linear interscale transfer}}
 +\\
 &\underbrace{\overline{  du_j'\frac{\partial dq^2}{\partial r_j}}}_{\text{Nonlinear interscale transfer}}
 =
 \underbrace{\overline{-2du_i'
  \frac{\partial
  dp'}{\partial X_i}
  }}_{\text{Pressure-velocity correlation}}
  -
  \underbrace{\overline{2du_i' u'^*_{j}\frac{\partial dU_i}{\partial X_j}}
 -
\overline{2du_i' du_j' \frac{\partial dU_i}{\partial r_j}} }_{\text{Production by mean flow } (=\mathcal{P})}\\
  &
  +
  \underbrace{\overline{\nu \frac{1}{2} \frac{\partial^2 dq^2}{\partial X_j \partial X_j}
}}_{\text{Physical diffusion}}
+ 
\underbrace{\overline{2\nu \frac{\partial^2 dq^2 }{\partial r_j \partial r_j}}}_{\text{Scale diffusion}}
-
\underbrace{
4\nu \left(  \frac{1}{4}
\overline{ \frac{\partial d u'_i}{\partial X_j} \frac{\partial d u'_i}{\partial X_j}}
+
\overline{\frac{\partial d u'_i}{\partial r_j} \frac{\partial d u'_i}{\partial r_j} } \right)}_{\text{Dissipation } (=\epsilon)}
\label{KHMH_standard}
\end{aligned}
\end{equation}

Note that $U^*_{j}$ and $u'^*_{j}$ denote the midpoint values of the time-average and fluctuating velocities respectively, i.e.\ $U^*_{j}=\left(U^+_j + U^-_j\right)/2$ and  $u'^*_{j}=\left(u'^+_j + u'^-_j\right)/2$. The physical meaning of each term is also provided; the production by mean flow and dissipation are denoted by $\mathcal{P}$ and $\epsilon$ respectively.  Assuming that the transient term is 0, the above equation can be written in divergence form as 
\begin{equation}
\frac{\partial \boldsymbol{\phi_s}_i
}{\partial X_i}
+
\frac{\partial \boldsymbol{\phi_r}_i}{\partial r_i} 
= 
\mathcal{P}-\epsilon,
\label{KHMH_3}
\end{equation}
where 
\begin{equation}
\boldsymbol{\phi_s}_i=\underbrace{\overline {U_i^*\delta q^2} }_{=\phi_{s_i}^M}
+
\underbrace{\overline{u'^*_i \delta q^2}}_{=\phi_{s_i}^F}
+
\underbrace{\overline{2 \delta u'_i \delta p'}}_{=\phi_{s_i}^P}
\underbrace{-\frac{1}{2}\nu \frac{\partial \overline {\delta q^2}}{\partial X_i}}_{=\phi_{s_i}^V},
\label{eq:flux_phys_space}
\end{equation}
is the total flux vector in physical space and
\begin{equation}
\boldsymbol{\phi_r}_i=
\underbrace{\overline {\delta u'_i \delta q^2}}_{=\phi_{r_i}^F}
+
\underbrace{\overline{\delta U_i \delta q^2}}_{=\phi_{r_i}^M}
\underbrace{-2\nu \frac{\partial \overline {\delta q^2}}{\partial r_i}}_{=\phi_{r_i}^V},
\label{eq:flux_scale_space}
\end{equation}
is the total flux vector in scale space. We use the superscript 'F' to denote the non-linear components,  $\phi_{s_i}^F$ and $\phi_{r_i}^F$, of these vectors respectively. The conditional decomposition of the non-linear fluxes $\phi_{s_i}^F$ and $\phi_{r_i}^F$ will be examined in section \ref{sec:cond_decompostition}. We are now ready to proceed with the derivation of the conditionally-averaged KHMH equation. 

\section{Conditionally-averaged KHMH equation}
\label{sec:conditionally_KHMH_equation}

\subsection{Definitions}\label{sec:conditional_operations}
In order to derive the conditionally-averaged form of the KHMH equation, we need first to define the conditions under which the averaging is performed. To this end, we employ the local binary function $\tau(X_i,t)$ to distinguish between instantaneous laminar and turbulent states at point $X_i$ at time $t$. More specifically, $\tau(X_i,t)$ takes the value of $0$ for the former (i.e.\ laminar), and $1$ for the latter (i.e.\ turbulent) state. This binary function is computed using the standard deviation of $D=|v|+|w|$;  refer to \cite{marxen2019turbulence} and \cite{Yao_Portela_Papadakis_2020} for more details.

Since the KHMH equation involves two points, the conditions for the averaging operation should be defined using the states at both points. Four  combinations are possible, 
\begin{enumerate}
    \item If both $x_i^+$ and $x_i^-$ are located within a turbulent patch, this is denoted as a  Turbulent-Turbulent (or $TT$) event, and it is defined by the condition $\tau^+\tau^-=1$, where $\tau^+=\tau(x_i^+,t)$ and $\tau^-=\tau(x_i^-,t)$.
    \item If both points are within the laminar region,  this is a  Laminar-Laminar (or $LL$) event, and it is defined by the condition $(1-\tau^+)(1-\tau^-)=1$.
    \item  If $x_i^+$, $x_i^-$  are within a turbulent and a laminar patch respectively, this is a Turbulent-Laminar (or $TL$) event, defined by $\tau^+(1-\tau^-)=1$.
    \item  If $x_i^+$, $x_i^-$  are within a laminar and turbulent region respectively, this is a Laminar-Turbulent (or $LT$) event, defined by $(1-\tau^+)\tau^-=1$.
\end{enumerate}

\begin{figure}
\centering
\includegraphics[scale=0.5]{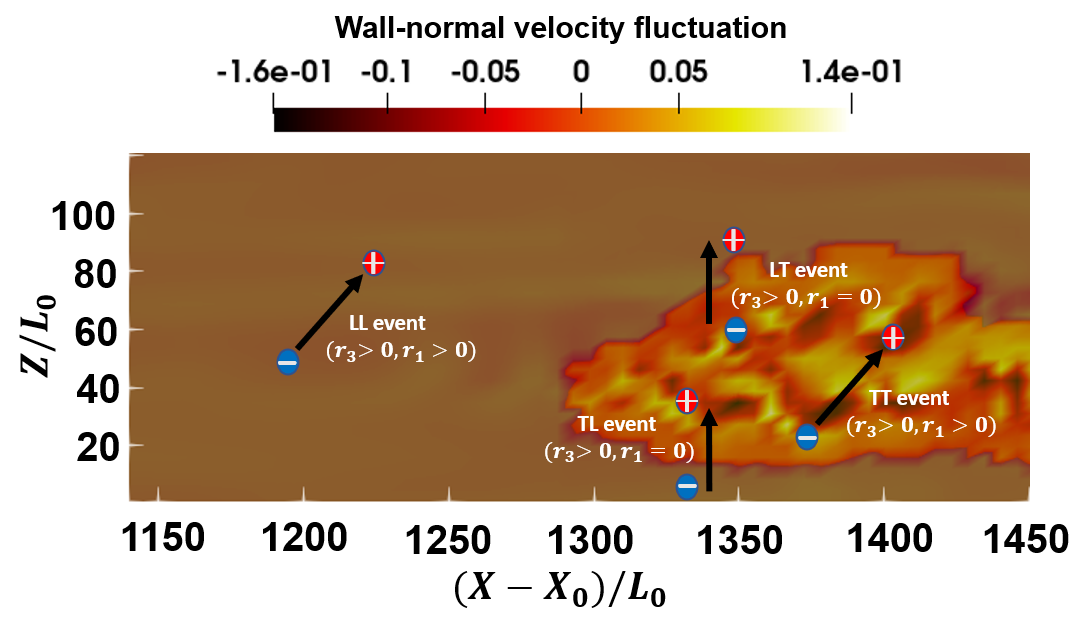}
\caption{Contour plot of wall-normal velocity fluctuations in the transitional region. A turbulent spot is clearly visualised on the right half of the figure. Within the spot $\tau=1$, while outside (grey region) $\tau=0$. The four different two-point event types, $LL$, $TT$, $TL$ and $LT$, are shown. 
}
\label{sche_r3}
\end{figure}

In our notation, the first capital letter denotes the state of point $x_i^+$ and the second the state of  $x_i^-$. We will also use the generic notation $AA$ to refer to a general event, i.e.\ $AA=TT$ or $LL$ or $TL$ or $LT$. The four events are shown schematically in figure \ref{sche_r3}. 

We can now define the time- and spanwise-average two-point intermittencies as, 
\begin{subequations}
 \begin{align}
\gamma^{(TT)}& \equiv \frac{1}{\Delta T L_z} \int_0^{\Delta T} \int_0^{L_z} \tau^{+}\tau^{-}  dz dt, \label{eq:intermittency_TT}
\\
\gamma^{(TL)}& \equiv \frac{1}{\Delta T L_z } \int_0^{\Delta T} \int_0^{L_z} \tau^{+}(1-\tau^{-})  dz dt, 
\\
\gamma^{(LT)}&\equiv \frac{1}{\Delta T L_z } \int_0^{\Delta T} \int_0^{L_z} (1-\tau^{+})\tau^{-}  dz dt,
\\
\gamma^{(LL)}&\equiv \frac{1}{\Delta T L_z } \int_0^{\Delta T} \int_0^{L_z} (1-\tau^{+})(1-\tau^{-}) dz dt. 
\end{align}
\label{gamma_equ}
\end{subequations}
Since $\tau^+\tau^-+(1-\tau^+)(1-\tau^-)+\tau^+(1-\tau^-)+(1-\tau^+)\tau^-=1$, we have $\gamma^{(TT)}+\gamma^{(TL)}+\gamma^{(LT)}+\gamma^{(LL)}=1$. Two point intermittencies were also defined in \cite{Yao_Portela_Papadakis_2020}, where $TL$ and $TL$ events were amalgamated as a combined $TL$ event. Here we consider the two event types separately for reasons that will become clear shortly.

The conditional time-averages of the general two-point variable, $dQ=Q(x_i^+)-Q(x_i^-)$, are defined as,
\begin{subequations}
\begin{align}
\overline{dQ}^{(TT)}(X,Y;r_1,r_3)&\equiv \frac{1}{\Delta T L_z \gamma^{(TT)}} \int_0^{\Delta T} \int_0^{L_z} \tau^{+}\tau^{-} dQ \: dz dt, 
\\
\overline{dQ}^{(TL)}(X,Y;r_1,r_3)&\equiv \frac{1}{\Delta T L_z \gamma^{(TL)}} \int_0^{\Delta T} \int_0^{L_z} \tau^{+}(1-\tau^{-}) dQ \: dz dt, 
\\
\overline{dQ}^{(LT)}(X,Y;r_1,r_3)&\equiv \frac{1}{\Delta T L_z \gamma^{(LT)}} \int_0^{\Delta T} \int_0^{L_z} (1-\tau^{+})\tau^{-} dQ \: dz dt,
\\
\overline{dQ}^{(LL)}(X,Y;r_1,r_3)&\equiv \frac{1}{\Delta T L_z \gamma^{(LL)}} \int_0^{\Delta T} \int_0^{L_z} (1-\tau^{+})(1-\tau^{-}) dQ \: dz dt. 
\end{align}
\label{calc_conperperties}
\end{subequations}
This means that the standard time-average can be decomposed as
\begin{equation}
 \begin{aligned}
\overline{dQ}(X,Y;r_1,r_3)  & =
\frac{1}{\Delta T L_z} \int_0^{\Delta T} \int_0^{L_z} dQ \: dz dt \\
& = \gamma^{(TT)} \overline{dQ}^{(TT)}+\gamma^{(TL)}\overline{dQ}^{(TL)}+
\gamma^{(LT)} \overline{dQ}^{(LT)}+\gamma^{(LL)} \overline{dQ}^{(LL)}
\end{aligned}
\label{eq:decomposition_linear}
\end{equation}

The definitions are similar for the conditionally average midpoint variable $Q^*=0.5\left[Q(x_i^+)+Q(x_i^-)\right]$ thus,
\begin{equation}
 \begin{aligned}
\overline{Q^*}(X,Y;r_1,r_3)  & =
\frac{1}{\Delta T L_z} \int_0^{\Delta T} \int_0^{L_z} Q^* \: dz dt \\
& = \gamma^{(TT)} \overline{Q^*}^{(TT)}+\gamma^{(TL)}\overline{Q^*}^{(TL)}+
\gamma^{(LT)} \overline{Q^*}^{(LT)}+\gamma^{(LL)} \overline{Q^*}^{(LL)}
\end{aligned}
\label{eq:decomposition_linear_Qstar}
\end{equation}

\begin{figure}
    \centering
  \includegraphics[scale=0.5]{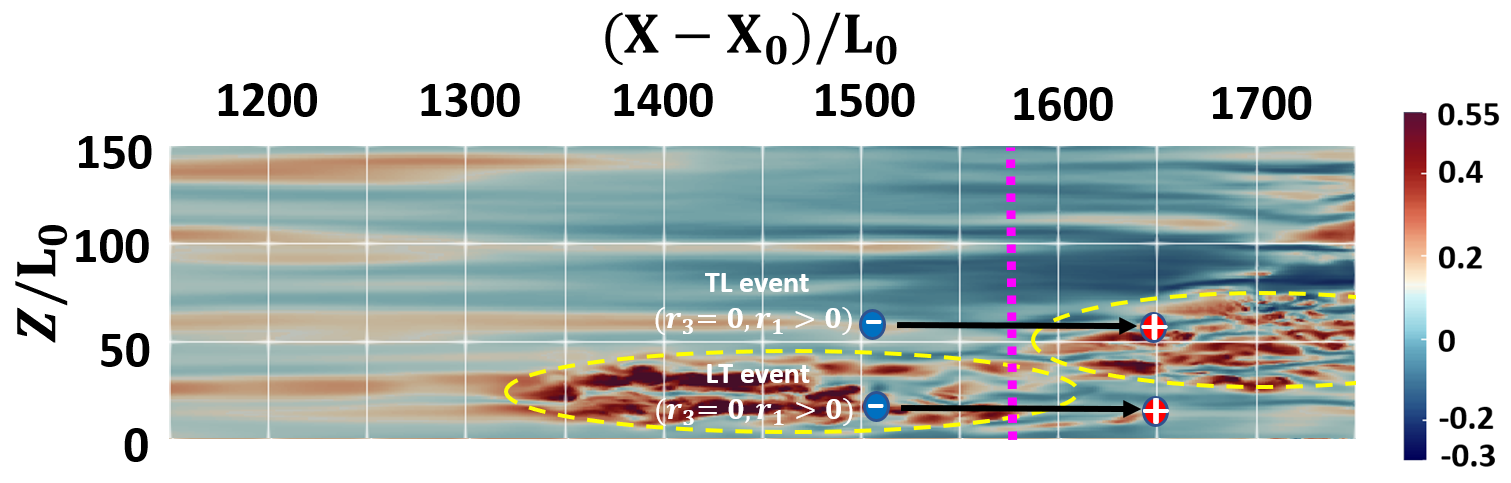}
    \caption{Contour plot of instantaneous streamwise velocity fluctuations. Yellow ovals demarcate two turbulent spots. The purple vertical dotted line represents a fixed streamwise location, X. Two event types, TL and LT, are shown with $r_1 \neq 0 $ and $r_3 = 0$.}
    \label{sche_r1}
\end{figure}

Referring back to figure \ref{sche_r3}, it is clear that $\overline{d Q}^{(TL)}$ and $\overline{d Q}^{(LT)}$ are two-point averages across the laminar/turbulent interface. For $r_1=0$ and $r_3\ne0$ (case shown in figure \ref{sche_r3}), these two conditional averages are taken across the interface in the spanwise direction. If ${dQ}$ is a quadratic function (for example $dQ=\overline{dq^2}$) due to homogeneity in $Z$ we have $\overline{d q^2}^{(TL)}(X,Y;0,r_3)=\overline{d q^2}^{(LT)}(X,Y;0,r_3)$. For $r_1\ne0$ and $r_3=0$ (case shown in figure \ref{sche_r1}), $LT$ averaging is taken across the tail (i.e.\ the downstream end) of a turbulent spot, while $TL$ is taken across the head of the spot (i.e.\ the upstream end). It is therefore possible to distinguish the different properties of the head or tail of a spot using the appropriate conditionally averaged quantity. This is an important observation and facilitates the physical interpretation of the results presented in section \ref{sec:evolution_cond_averaged_fluxes}.   

We can now proceed to derive the conditionally-averaged KHMH equation.  

\subsection{Derivation of the conditionally-averaged KHMH equation}
We start again with the Navier-Stokes equations at two points $x_i^+$ and $x_i^-$, equations \eqref{eq:momentum_at_x+_and_x-}, and define the  conditional velocity fluctuation difference as
\begin{equation}
du_i'^{(AA)} \equiv du_i-dU_i^{(AA)},
\label{BB1}
\end{equation}
\noindent where $AA =TT$ or $LL$ or $TL$ or $LT$ as mentioned earlier,  $du_i=u_i^+-u_i^-$ and $dU_i^{(AA)}=\overline{du_i}^{(AA)}$ (from definition  \eqref{calc_conperperties}). It is straightforward to prove that  $\overline{du_i'^{(AA)}}^{(AA)}=0$; this is the equivalent of $\overline{du_i'}=0$ in standard averaging. 

Similarly, the conditional fluctuation velocity at the midpoint is defined as 
\begin{equation}
{u'^*_i}^{(AA)} \equiv {u^*_i}-{U^*_i}^{(AA)},
\label{BB2}
\end{equation}
where $u^*_{i}=0.5\left(u_i^++u_i^-\right)$ and ${U^*_i}^{(AA)}=\overline{u^*_{i}}^{(AA)}$, and again
$\overline{{u'^*_i}^{(AA)}}^{(AA)}=0$. 

Subtracting equation \eqref{eq:momentum_at_x-} from \eqref{eq:momentum_at_x+}, multiplying each term by $2du_i'^{(AA)}$, and then applying the $(AA)$ averaging operation as defined in equation \eqref{calc_conperperties} we obtain,
 \begin{equation}
 \begin{aligned}
  \underbrace{
  \overline{2du_i'^{(AA)}\frac{\partial du_i}{\partial t}}^{(AA)}}_{\text{Transient term}}
 + 
  \underbrace{
  \overline{2du_i'^{(AA)}u^+_j \frac{\partial du_i}{\partial x^+_j} 
 +
 2du_i'^{(AA)}u^-_j \frac{\partial du_i}{\partial x^-_j}}^{(AA)}}_{\text{Non-linear term}}
 = \\
  \underbrace{
   \overline{-2du_i'^{(AA)}(\frac{\partial dp}{\partial x^+_i} 
 -
 \frac{\partial dp}{\partial x^-_i})}^{(AA)} }_{\text{Pressure-velocity correlation}}
 +
 \underbrace{
  \overline{2 \nu du_i'^{(AA)}\frac{\partial^2 du_i}{\partial x_j^+ \partial x_j^+} 
+
2\nu du_i'^{(AA)}\frac{\partial^2 du_i}{\partial x_j^- \partial x_j^-}}^{(AA)}
}_{\text{Viscous term}},
\label{Fin2}
\end{aligned}
\end{equation}
where we have used again $\dfrac{\partial u^+_i}{\partial x^-_j}=0$ and $\dfrac{\partial u^-_i}{\partial x^+_j}=0$. This is the equivalent of \eqref{eq:standard_time_average_Fin2} for standard-averaging.

We now define the two-point conditional energy as $\overline{d q^{2^{(AA)}}}^{(AA)} =\overline{d u_i'^{(AA)} d u_i'^{(AA)}}^{(AA)}$ and seek its transport equation in the physical and scale spaces, similar to \eqref{KHMH_standard}. Applying again the variable transformation $X_i=0.5\left(x^+_i + x^-_i\right)$ and  $r_i = x^+_i - x^-_i$ and the definitions \eqref{BB1} and \eqref{BB2} into \eqref{Fin2}, after some algebra we  get the following conditionally-averaged KHMH equation for $\overline{d q^{2^{(AA)}}}^{(AA)}$,
 \begin{equation}
 \begin{aligned}
& \underbrace{\overline{\frac{\partial  dq^{2^{(AA)}}}{\partial t}
 }^{(AA)}}_{\text{Transient term}}
 + 
 \underbrace{\overline{ {U^*_j}^{(AA)}\frac{\partial q^{2^{(AA)}}}{\partial X_j} }^{(AA)}}_{\text{Mean advection}}
 +
  \underbrace{\overline{ {u'^*_j}^{(AA)}\frac{\partial dq^{2^{(AA)}}}{\partial X_j} }^{(AA)}}_{\text{Turbulent advection}}
+
 \underbrace{\overline{ dU_j^{(AA)} \frac{\partial dq^{2^{(AA)}}}{\partial r_j} }^{(AA)}}_{\text{Linear transfer}}\\
 &+
 \underbrace{\overline{ du_j'^{(AA)} \frac{\partial dq^{2^{(AA)}}}{\partial r_j}}^{(AA)} }_{\text{Nonlinear transfer}}
 =
 \underbrace{\overline{-2du_i'^{(AA)}
  \frac{\partial
  dp'^{(AA)}}{\partial X_i}
  }^{(AA)}}_{\text{Pressure-velocity correlation}}
    +
  \underbrace{\nu \frac{1}{2} \overline{ \frac{\partial^2 dq^{2^{(AA)}}}{\partial X_j \partial X_j}
}^{(AA)}}_{\text{Physical diffusion}}
  + 
\underbrace{2\nu \overline{ \frac{\partial^2 dq^{2^{(AA)}} }{\partial r_j \partial r_j}}^{(AA)}}_{\text{Scale diffusion}}\\
  &
  \underbrace{-\overline{2du_i'^{(AA)} {u'^*_j}^{(AA)}\frac{\partial dU_i^{(AA)}}{\partial X_j}}^{(AA)}
 -
\overline{2du_i'^{(AA)} du_j'^{(AA)} \frac{\partial dU_i^{(AA)}}{\partial r_j}}^{(AA)} }_{\text{Production}}\\
  &
-
\underbrace{2\nu \overline{ \left( \frac{\partial du'^{(AA)}_i }{\partial x^+_j}\frac{\partial du'^{(AA)}_i }{\partial x^+_j}
+ 
\frac{\partial du'^{(AA)}_i }{\partial x^-_j}\frac{\partial du'^{(AA)}_i }{\partial x^-_j} \right ) }^{(AA)}}_{\text{Dissipation}}
\label{KH_CON}
\end{aligned}
\end{equation}

The form of equation \eqref{KH_CON} is similar to the standard form  \eqref{KHMH_standard}. This is due to the appropriate  definitions of the conditional fluctuating quantities in \eqref{BB1} and \eqref{BB2} that satisfy   $\overline{du_i'^{(AA)}}^{(AA)}=0$ and $\overline{{u'^*_i}^{(AA)}}^{(AA)}=0$. There is however an important difference. Standard averaging commutes with the spatial differentiation operation, for example,
\begin{equation}
\frac{\partial \overline{d Q}}{\partial r_j}= \frac{1}{\Delta T L_z}
\frac{\partial}{\partial r_j} \left( \int_0^{\Delta T} \int_0^{L_z} d Q dz dt \right)=
\frac{1}{\Delta T L_z}
 \int_0^{\Delta T} \int_0^{L_z} \frac{\partial d Q}{\partial r_j}  dz dt =\overline{\frac{\partial {d Q}}{\partial r_j}}
\end{equation}
This is however not the case for conditionally-averaged quantities, for example
\begin{equation}
\frac{\partial \overline{d Q}^{(TT)}}{\partial r_j}= \frac{1}{\Delta T L_z}
\frac{\partial}{\partial r_j} \left( \frac{1}{\gamma^{(TT)}} \int_0^{\Delta T} \int_0^{L_z} \tau^{+}\tau^{-} d Q dz dt \right),
\end{equation}
while
\begin{equation}
\overline{\frac{\partial {d Q}}{\partial r_j}}^{(TT)}= \frac{1}{\Delta T L_z}
\left( \frac{1}{\gamma^{(TT)}} \int_0^{\Delta T} \int_0^{L_z} \tau^{+}\tau^{-} \frac{\partial {d Q}}{\partial r_j} dz dt \right) \ne \frac{\partial \overline{d Q}^{(TT)}}{\partial r_j},
\end{equation}
\noindent because $\gamma^{(TT)}$ depends on $r_j$. The same issue appears in the conditionally-averaged TKE equation \citep{marxen2019turbulence}. 

This lack of commutation has two important implications. First, equation \eqref{KH_CON} cannot be written in conservative form. For example, for the non-linear inter-scale energy transfer term, we have
 \begin{equation}
\overline{ du_j'^{(AA)} \frac{\partial dq^{2^{(AA)}}} {\partial r_j}}^{(AA)} 
=
\overline{ \frac{\partial dq^{2^{(AA)}} du_j'^{(AA)}}{\partial r_j}}^{(AA)}  
\neq
 \frac{\partial \left(\overline{dq^{2^{(AA)}} du_j'^{(AA)}}^{(AA)} \right)}{\partial r_j}
 \label{eq:non-commutation}
\end{equation}
The first equality in the above equation is because $\dfrac{\partial{du_j'^{(AA)}}}{\partial r_j}=0$ (this is  equivalent to $\dfrac{\partial \overline{du_j'}}{\partial r_j}=0$ for standard averaging). It is easy to prove; for example if $AA=TT$, we have
\begin{equation}
\begin{aligned}
&\frac{\partial du_j'^{(TT)}}{\partial r_j}= 
\frac{\partial du_j}{\partial r_j}-\frac{\partial dU_j^{(TT)}}{\partial r_j}=\frac{\partial du_j}{\partial r_j}-
\frac{\partial}{\partial r_j} \left( \frac{1}{{\Delta T L_z} \gamma^{(TT)}} \int_0^{\Delta T} \int_0^{L_z} \tau^{+}\tau^{-} dU_j  dz dt \right)=\\
&\frac{\partial du_j}{\partial r_j}-
\frac{\partial}{\partial r_j} \left( \frac{dU_j }{\Delta T L_z \gamma^{(TT)}} \int_0^{\Delta T} \int_0^{L_z} \tau^{+}\tau^{-}  dz dt \right)=
\frac{\partial du_j}{\partial r_j}-
\frac{\partial dU_j}{\partial r_j}=0-0=0,
\end{aligned}
\label{ineq}
\end{equation}
because $du_j$ and $dU_j$ satisfy the continuity equation in the scale space. We also took into account that $dU_j$ is constant with respect to $Z$ and $t$ and used \eqref{eq:intermittency_TT}. 

The second implication is computational. For standard averaging, all the terms in the KHMH equation can be evaluated numerically by differentiating locally at points $x_i^+$ and $x_i^-$, for example  
\begin{equation}
 \begin{aligned}
 \frac{\partial  dU_i }{\partial r_j}=
\frac{\partial  \overline{du_i} }{\partial r_j}
=
\overline{\frac{\partial {du_i} }{\partial r_j}}
&=
\frac{1}{2}\left ( \overline{\frac{\partial {(u_i^+ - u_i^-)} }{\partial x^+}}
-
\overline{\frac{\partial {(u_i^+ - u_i^-)} }{\partial x_j^-}}
\right )
=
\frac{1}{2} \left( \overline{\frac{\partial {u_i^+} }{\partial x_j^+}}
+
\overline{\frac{\partial  {u_i^-} }{\partial x_j^-}}
\right ) 
=\\
&\frac{1}{2} \left( \frac{\partial  \overline{u_i^+} }{\partial x_j^+}
+
\frac{\partial  \overline{u_i^-}}{\partial x_j^-}
\right )
=
\frac{1}{2} \left( {\frac{\partial U_i^+ }{\partial x_j^+}}
+
{\frac{\partial  U_i^- }{\partial x_j^-}}
\right ) 
\end{aligned}
\end{equation}

However, this is not the case for the conditionally-averaged velocity difference, 
\begin{equation}
 \begin{aligned}
\frac{\partial  dU_i^{(AA)} }{\partial r_j}=\frac{\partial  \overline{du_i}^{(AA)} }{\partial r_j}
\ne \overline{\frac{\partial {du_i} }{\partial r_j}}^{(AA)}
&
\left[ =
\frac{1}{2} \left ( \overline{\frac{\partial {(u_i^+ - u_i^-)}}{\partial x_j^+}}^{(AA)}
-
\overline{\frac{\partial {(u_i^+ - u_i^-)}}{\partial x_j^-}}^{(AA)}
\right ) \right. \\ 
&  =
\left .
\frac{1}{2} \left (\overline{ 
\frac{{\partial u_i^+ }}{\partial x_j^+}}^{(AA)}
+
\overline{ 
\frac{{\partial  u_i^-}}{\partial x_j^-}}^{(AA)}
\right ) \right] 
\label{diff}
\end{aligned}
\end{equation}
This means that the derivatives of the conditionally-averaged two-point variables must be calculated directly in scale space. The process and validation against standard averaging are presented in appendix \ref{sec:appendix_numerical_calculation}. 

\section{Conditional decomposition of non-linear energy fluxes}\label{sec:cond_decompostition}
In this section we decompose the non-linear energy fluxes in scale and physical spaces, $\phi_{r_j}^F=\overline{dq^2du'_j}=\overline{(du'_i)^2du'_j}$ and  ${\phi_{s_j}^F}=\overline {u'^*_j \delta q^2}$  respectively, see equations \eqref{eq:flux_scale_space} and \eqref{eq:flux_phys_space}, into conditionally-averaged components, i.e.\ we seek to derive expressions similar to \eqref{eq:decomposition_linear} and \eqref{eq:decomposition_linear_Qstar}. The decomposition for the second-order structure function (quadratic quantity) was derived in  \cite{Yao_Portela_Papadakis_2020}. Here we extend the method to energy fluxes (cubic quantities). 

For the non-linear energy flux in scale space $\phi_{r_j}^F=\overline{(du'_i)^2du'_j}$ we have,  

\begin{equation}
\begin{aligned}
\phi_{r_j}^F &=\overline{(du'_i)^2du'_j}=
\overline{(du_i-dU_i)^2(du_j-dU_j)}
\\
&=\overline{
(du_i)^2du_j}
-
\overline{(du_i)^2}dU_j
-\overline{2du_idu_j}dU_i
+2dU_idU_idU_j \\
&= 
\gamma^{(TT)}\overline{(du_i)^2du_j}^{(TT)}
+
\gamma^{(TL)}\overline{(du_i)^2du_j}^{(TL)}
+
\gamma^{(LT)}\overline{(du_i)^2du_j}^{(LT)}
+\\
& 
\gamma^{(LL)}\overline{(du_i)^2du_j}^{(LL)}
\quad-
\overline{(du_i)^2}dU_j
-\overline{2du_idu_j}dU_i
+2dU_idU_idU_j,
\end{aligned}
\label{final_phiF_before}
\end{equation}
where \eqref{eq:decomposition_linear} was applied to $\overline{(du_i)^2du_j}$. 

We now write $\overline{(du_i)^2du_j}^{(AA)}$ ($AA =TT$ or $LL$ or $TL$ or $LT$) in terms of conditional fluctuations. To do this, we use the definition $du_i'^{(AA)} \equiv du_i-dU_i^{(AA)}$, see equation \eqref{BB1}, and express the  conditional non-linear energy flux ${\phi^F_{r_j}}^{(AA)}=\overline{\left(du'^{(AA)}_i\right)^2du'^{(AA)}_j}^{(AA)}$ as 
\begin{equation}
 \begin{aligned}
{\phi^F_{r_j}}^{(AA)}=\overline{\left(du'^{(AA)}_i\right)^2du'^{(AA)}_j}^{(AA)}
&=
\overline{
(du_i)^2du_j}^{(AA)}
-
\overline{(du_i)^2}^{(AA)}dU_j^{(AA)} \\
&-\overline{2du_idu_j}^{(AA)}dU_i^{(AA)}
-2dU_i^{(AA)}dU_i^{(AA)}dU_j^{(AA)}
\end{aligned}
\label{phi_F_dcom}
\end{equation}
Solving for $\overline{(du_i)^2du_j}^{(AA)}$ and substituting into \eqref{final_phiF_before}, we obtain the desired decomposition
\begin{equation}
 \begin{aligned}
&\phi^F_{r_j}
=\gamma^{(TT)}{\phi^F_{r_j}}^{(TT)}+
\gamma^{(TL)}{\phi^F_{r_j}}^{(TL)}+
\gamma^{(LT)}{\phi^F_{r_j}}^{(LT)}+
\gamma^{(LL)}{\phi^F_{r_j}}^{(LL)}+\phi_{r_j},
\end{aligned}
\label{final_phiF}
\end{equation}
where the additional term $\phi_{r_j}$ is given by,
\begin{equation}
 \begin{aligned}
\phi_{r_j} &=
\gamma^{(TT)}\left[ \overline{(du_i)^2}^{(TT)}dU_j^{(TT)}
+\overline{2du_idu_j}^{(TT)}dU_i^{(TT)}
-2dU_i^{(TT)}dU_i^{(TT)}dU_j^{(TT)} \right]\\
&
+
\gamma^{(TL)} \left[
\overline{(du_i)^2}^{(TL)}dU_j^{(TL)}
+\overline{2du_idu_j}^{(TL)}dU_i^{(TL)}
-2dU_i^{(TL)}dU_i^{(TL)}dU_j^{(TL)} \right] \\
&
+
\gamma^{(LT)}\left[
\overline{(du_i)^2}^{(LT)}dU_j^{(LT)}
+\overline{2du_idu_j}^{(LT)}dU_i^{(LT)}
-2dU_i^{(LT)}dU_i^{(LT)}dU_j^{(LT)} \right] \\
&
+
\gamma^{(LL)}\left[
\overline{(du_i)^2}^{(LL)}dU_j^{(LL)}
+\overline{2du_idu_j}^{(LL)}dU_i^{(LL)}
-2dU_i^{(LL)}dU_i^{(LL)}dU_j^{(LL)} \right]\\
&-
\overline{(du_i)^2}dU_j
-\overline{2du_idu_j}dU_i
+2dU_idU_idU_j 
\end{aligned}
\label{final_phiF_PHI}
\end{equation}

\begin{figure}
    \centering
  \includegraphics[scale=0.414]{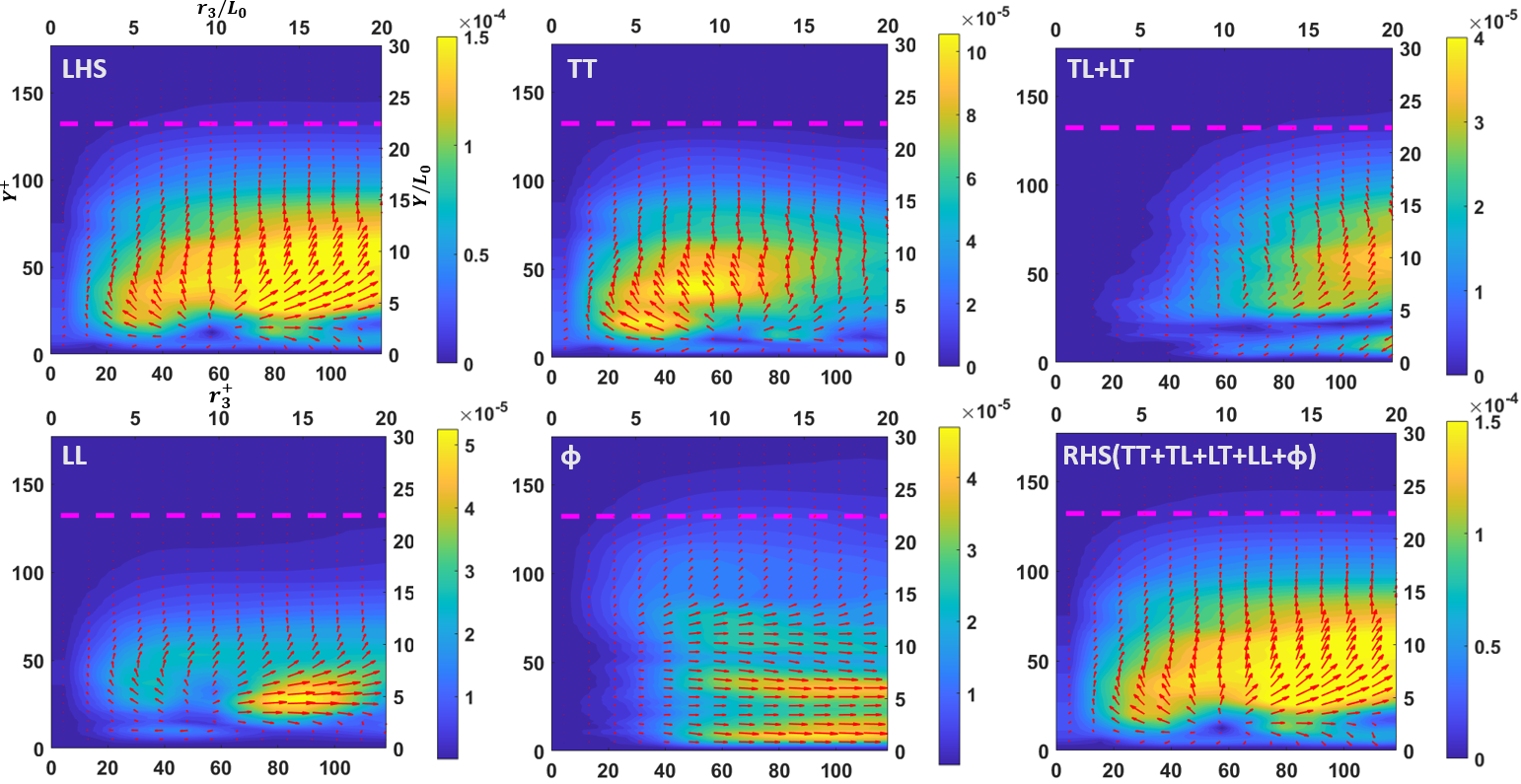}
    \caption{Non-linear interscale energy flux vector $\left(\phi^F_{r_3},\phi^F_{s_2}\right)$ (LHS) and the constituent components $\left(\gamma^{(AA)}{\phi^F_{r_3}}^{(AA)}, \gamma^{(AA)}{\phi^F_{s_2}}^{(AA)}\right)$ (where $AA$=$TT$,$TL+LT$,$LL$) in the $(r_3,Y)$ plane at location $TR2$. RHS denotes the sum of the terms on the right-hand side of equation \eqref{final_phiF}. Contours represent the magnitude of the flux vectors and the purple horizontal lines indicate the local boundary layer thickness. 
    }
    \label{Validation_decom}
\end{figure}

Following the same process, the following decomposition can be obtained for the non-linear energy flux in physical space,  ${\phi_{s_j}^F}=\overline {u'^*_j \delta q^2}=\overline {u'^*_j (du'_i)^2}$, 

\begin{equation}
 \begin{aligned}
&\phi^F_{s_j}
=\gamma^{(TT)}{\phi^F_{s_j}}^{(TT)}+
\gamma^{(TL)}{\phi^F_{s_j}}^{(TL)}+
\gamma^{(LT)}{\phi^F_{s_j}}^{(LT)}+
\gamma^{(LL)}{\phi^F_{s_j}}^{(LL)}+\phi_{s_j},
\end{aligned}
\label{final_phiF_S}
\end{equation}
where  ${\phi^F_{s_j}}^{(AA)}=\overline{{u'^*_j}^{(AA)}\left(du'^{(AA)}_i\right)^2}^{(AA)}$ and the additional term $\phi_{s_j}$ is given by,

\begin{equation}
 \begin{aligned}
\phi_{s_j}&=
\gamma^{(TT)}\left[\overline{(du_i)^2}^{(TT)}{U^*_j}^{(TT)}
+\overline{2du_iu^*_{j}}^{(TT)}dU_i^{(TT)}
-2dU_i^{(TT)}dU_i^{(TT)}{U^*_j}^{(TT)} \right]\\
&
+
\gamma^{(TL)}\left[]
\overline{(du_i)^2}^{(TL)}{U^*_j}^{(TL)}
+\overline{2du_iu^*_{j}}^{(TL)}dU_i^{(TL)}
-2dU_i^{(TL)}dU_i^{(TL)}{U^*_j}^{(TL)} \right] \\
&
+
\gamma^{(LT)}\left[
\overline{(du_i)^2}^{(LT)}{U^*_j}^{(LT)}
+\overline{2du_iu^*_{j}}^{(LT)}dU_i^{(LT)}
-2dU_i^{(LT)}dU_i^{(LT)}{U^*_j}^{(LT)} \right] \\
&
+
\gamma^{(LL)}\left[
\overline{(du_i)^2}^{(LL)}{U^*_j}^{(LL)}
+\overline{2du_iu^*_{j}}^{(LL)}dU_i^{(LL)}
-2dU_i^{(LL)}dU_i^{(LL)}{U^*_j}^{(LL)} \right]\\
&-
\overline{(du_i)^2}U^*_j
-\overline{2du_iu^*_{j}}dU_i
+2dU_idU_iU^*_j \\
\end{aligned}
\end{equation}

Figure \ref{Validation_decom} presents contours of all the terms appearing in \eqref{final_phiF} for the  decomposition of the energy flux vector $(\phi^F_{r_3},\phi^F_{s_2})$ at location $TR2$ on $(r_3, Y)$ plane (the separations in the other two directions are equal to 0, i.e.\ $r_1=0, r_2=0$).  In the figure, we combine the results of $TL$ and $LT$ together due to homogeneity in the spanwise direction. The upper left panel, denoted as LHS, depicts the results from the direct calculation of the flux vector $(\phi^F_{r_3},\phi^F_{s_2})$  using standard time-average, i.e.\ $\left(\overline{\left(du'_i\right)^2du'_3}, \overline{\left(du'_i\right)^2u'^*_{2}}\right)$. The lower right panel, denoted as RHS, presents the sum of all terms in the \eqref{final_phiF}. It is clear that the results in the two panels are essentially identical, confirming the validity of the derived decomposition. The flux vector originates from the focal point $Y/L_0 \approx 2.5$ and $r_3/L_0 \approx 10$ and transfers energy radially to different directions. The strongest flux is located in the region $0<Y/L_0<10, r_3/L_0>10$ and is found to be positive, indicating strong inverse cascade. This pattern was observed in \cite{yao2022analysis}. 

Apart from verifying the decomposition  \eqref{final_phiF}, figure \ref{Validation_decom} provides important insight into the  origin of the aforementioned energy flux pattern. It is interesting to observe that the inverse cascade arises from the $TL+LT$, $LL$, and $\phi$ components; all contribute to the strong positive flux in the region $0<Y/L_0<10$,  $r_3/L_0>10$ with values that are of the same order of magnitude. Outside this region, they have small values. On the other hand, the $TT$ component contributes to the forward cascade in the region left of the focal point. The shape therefore of the energy flux vectors on the $(r_3, Y)$ plane arises from the superposition of the contribution of $TT$ term which is responsible for forward cascade, and all the other terms which are responsible for inverse cascade. 

The constituent terms plotted in figure \ref{Validation_decom} are weighted by the corresponding two-point intermittencies, $\gamma^{(AA)}$, which can have very small values depending on the location examined,  refer to figure \ref{Cf_sch}. In the following section, the evolution of the vectors $\left( {{\phi^F_{r_3}}^{(AA)}},{{\phi^F_{s_2}}^{(AA)}} \right )=\left(\overline{\left(du'^{(AA)}_i\right)^2du'^{(AA)}_3}^{(AA)}, \overline{\left(du'^{(AA)}_i\right)^2{u'^*_2}^{(AA)}}^{(AA)}\right)$ at different streamwise locations is examined. 

\section{Evolution of conditionally-averaged non-linear energy fluxes} \label{sec:evolution_cond_averaged_fluxes}
\subsection{Fluxes on $(r_3,Y)$ plane}

\begin{figure}
    \centering
  \includegraphics[scale=0.372]{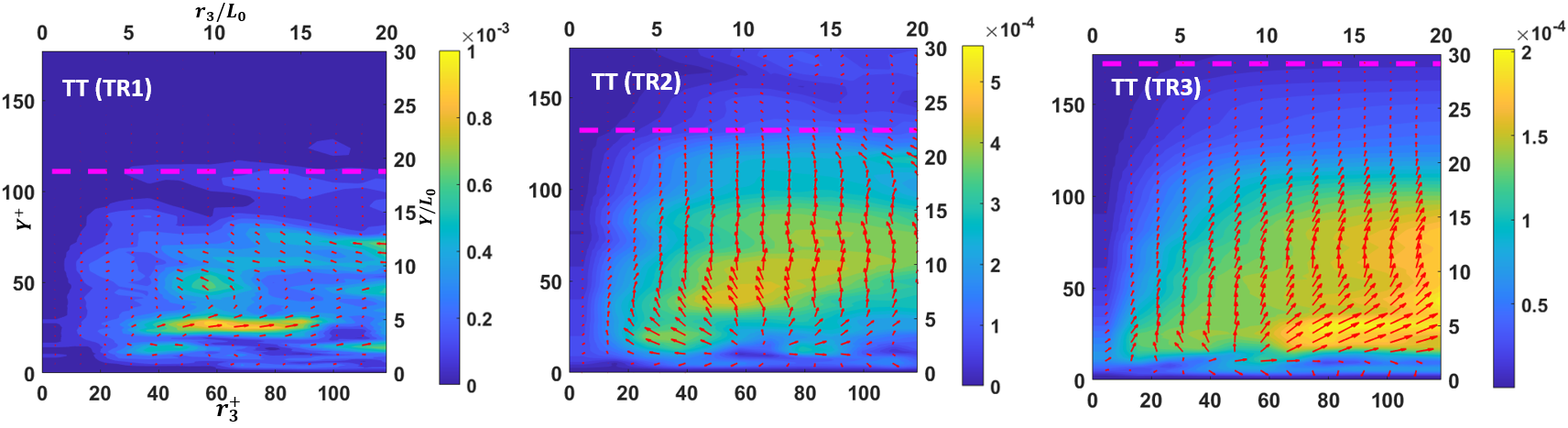}
  \includegraphics[scale=0.372]{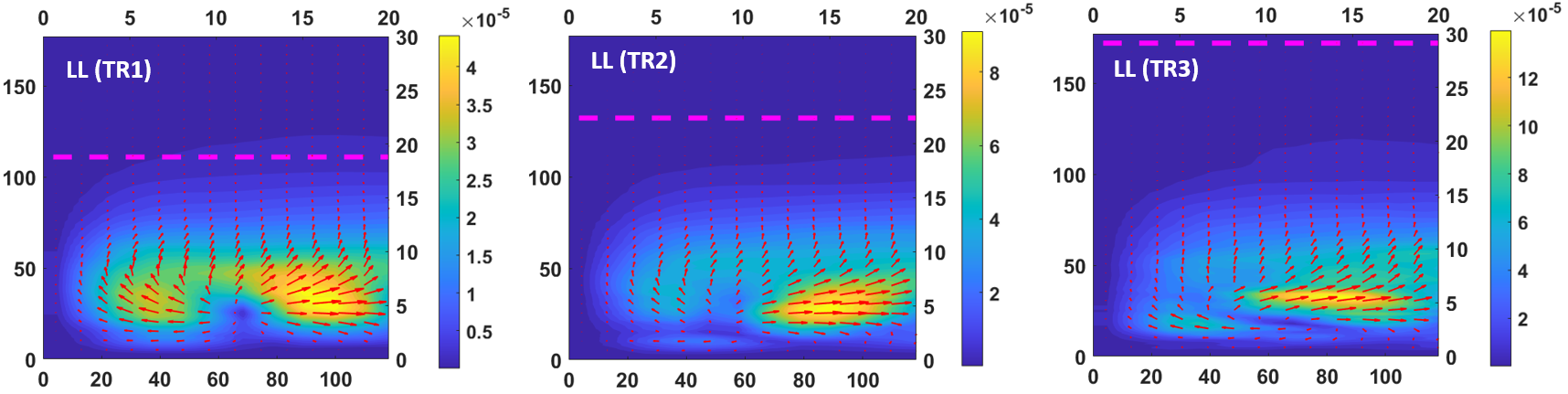}
   \includegraphics[scale=0.373]{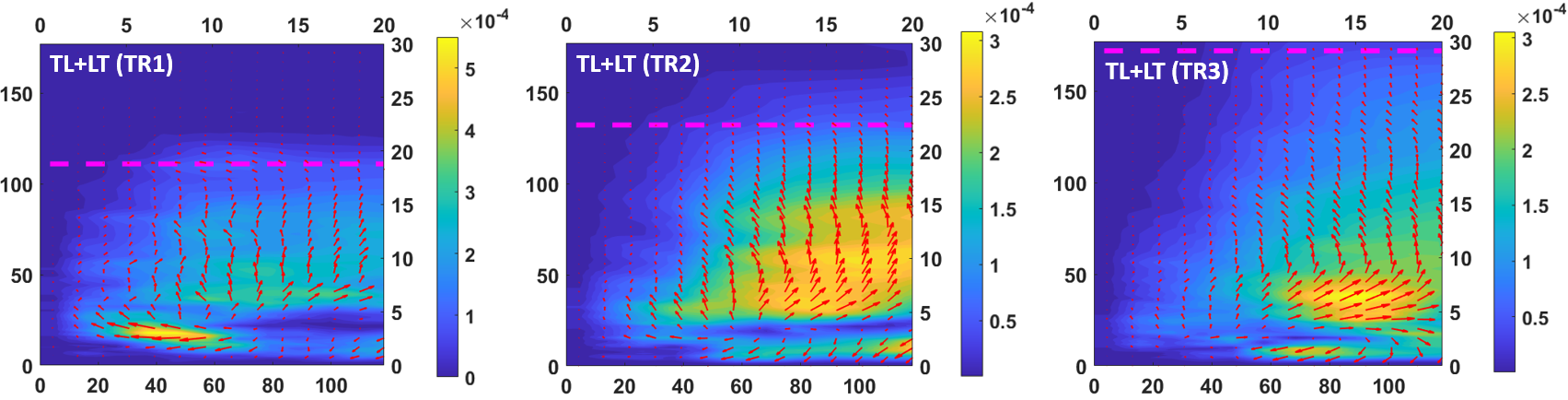}
    \includegraphics[scale=0.372]{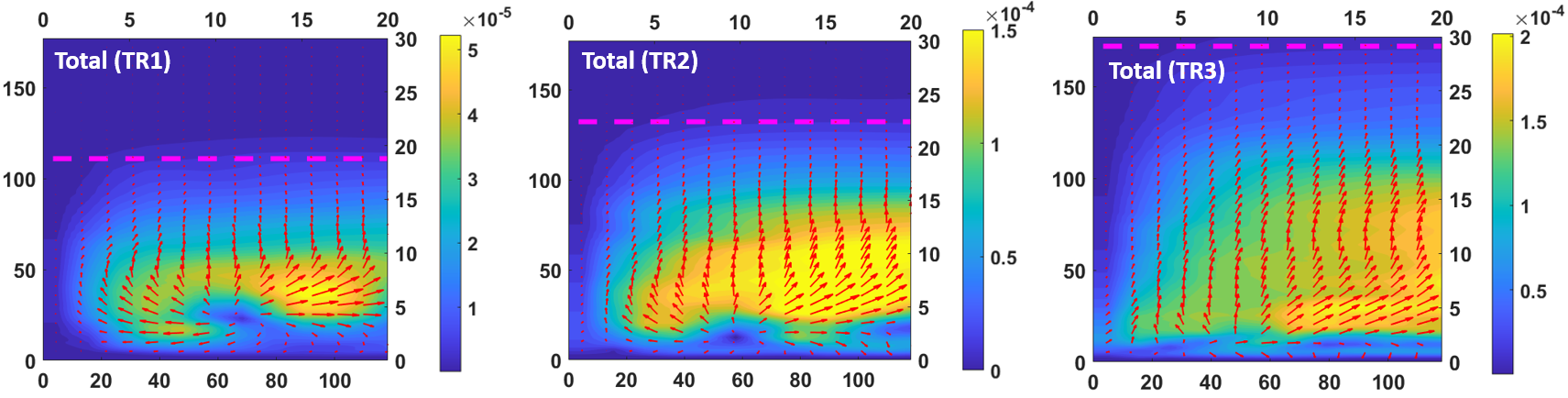}
    \caption{Conditionally-averaged flux vectors $\left( {{\phi^F_{r_3}}^{(AA)}},{{\phi^F_{s_2}}^{(AA)}} \right )$ (first, second and third row respectively starting from the top) and standard-averaged flux vector $\left( {{\phi^F_{r_3}}},{{\phi^F_{s_2}}} \right )$ (bottom row) at locations $TR1$ (left column), $TR2$ (middle column) and $TR3$ (right column). The contours represent the magnitude of the corresponding flux vectors and the purple horizontal lines indicate the local boundary layer thickness. 
    }
    \label{noweight_TR123}
\end{figure}

\begin{figure}
    \centering
  \includegraphics[scale=0.3]{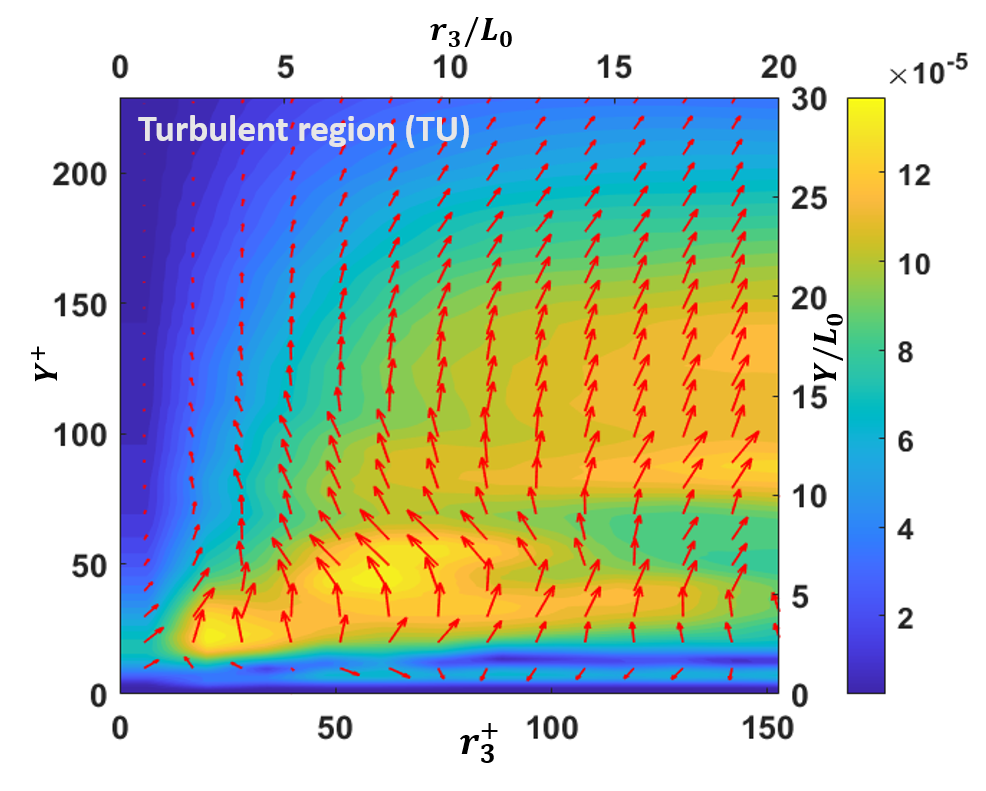}
    \caption{Standard-averaged flux vector $\left( {{\phi^F_{r_3}}},{{\phi^F_{s_2}}} \right )$ at $TU$ location. The contour represents the magnitude of the flux vectors.}
    \label{noweight_TU}
\end{figure}

Figure \ref{noweight_TR123} shows the conditionally averaged energy fluxes $\left( {{\phi^F_{r_3}}^{(AA)}},{{\phi^F_{s_2}}^{(AA)}} \right )$ where $AA$$=$ $TT$ or $TL+LT$ or $LL$ at locations $TR1,TR2,TR3$. The last row (marked as 'Total') depicts $\left( {{\phi^F_{r_3}}},{{\phi^F_{s_2}}} \right )$. In all plots, $r_1=0$ and $r_2=0$.

At the early stages of transition, at location $TR1$ (left column), the flux vector $\left( {{\phi^F_{r_3}}^{(LL)}},{{\phi^F_{s_2}}^{(LL)}} \right )$ is dominant and is almost identical to the total flux.  We can clearly see strong inverse cascade occurring to the right of the focal point, which is mainly due to $LL$ events.  The vectors $\left( {{\phi^F_{r_3}}^{(TT)}}, {{\phi^F_{s_2}}^{(TT)}} \right )$  and $\left( {{\phi^F_{r_3}}^{(TL)}}, {{\phi^F_{s_2}}^{(TL)}} \right )+\left( {{\phi^F_{r_3}}^{(LT)}},{{\phi^F_{s_2}}^{(LT)}} \right )$ are very localised and their behaviour is difficult to interpret. This may be due to the small number of $TT$ and $TL/LT$ events at this early transition location. Notice however the high values of the magnitude of the $TT$ vector, $\left({{\phi^F_{r_3}}^{(TT)}}, {{\phi^F_{s_2}}^{(TT)}} \right )$, compared to all the other vectors. However, its overall contribution to the total flux, $\gamma^{(TT)}\boldsymbol{\phi^F}^{(TT)}$, is small because the values of intermittency $\gamma^{(TT)}$ are negligible at this location (see figure \ref{Cf_sch}).

At location $TR2$ (middle column), $TT$ and $TL/LT$ events start to play a more significant role as expected, but they act in different areas of the map.  Events of the $LL$ type maintain their dominant contribution to the inverse cascade, while $TL$ \& $LT$ events also amplify and contribute to the inverse cascade in the area $10L_0<r_3<20L_0$ above $Y\approx 5L_0$. Notice again that $TT$-averaged flux vectors have a much higher magnitude compared to the total flux, and show a mixture of the forward and inverse cascade to the left and right of the focal point respectively (with the forward cascade being slightly stronger). This picture is consistent with figure \ref{Validation_decom} where the conditional-averaged fluxes are weighted by the two-point intermittency.

At location $TR3$ (right column), $TT$ events now assume the dominant role because $\gamma^{(TT)}$ approaches 1, $LL$ events are localised (in the same way that $TT$ events were localised at $TR1$ location), while $TL$ \& $LT$ events again give rise to inverse cascade. Interestingly, as intermittency increases, $TT$ events show stronger inverse cascade.  For comparison, the flux vector  $\left( \phi^F_{r_3}, \phi^F_{s_2} \right )$  at the fully turbulent location $TU$ is plotted in figure \ref{noweight_TU}. At this location, the inverse cascade is weakened (but it is still visible) and energy flows in the wall-normal direction before looping back to small spanwise length scales.   

The strong inverse cascade found at $TR2$ and $TR3$ locations is clearly not observed in the fully turbulent region. The origin of the inverse cascade arises mainly due to $TL$ \& $LT$ events. Indeed, in the $TR2$ location their magnitude is 5-6 times larger compared to $LL$ events (that also contribute to inverse cascade), while in $TR3$ about 2-3 times larger. The laminar/turbulent interface, therefore, plays a crucial role in the inverse cascade process. In the next section, we focus on the $TR2$ location and explore in more depth the cascade process in the three-dimensional $(r_1,r_3, Y)$ hyperplane. 

\subsection{Fluxes on the $(r_1,r_3,Y)$ hyperplane and  $(r_1,r_3)$ plane}

We project the conditionally-averaged fluxes ${\phi^F}^{(AA)}$ and ${\phi^S}^{(AA)}$ on the  $(r_1,r_3,Y)$ hyperplane, i.e.\ plot  three-dimensional maps of the vector $\left({\phi^F_{r_1}}^{(AA)}, {\phi^F_{r_3}}^{(AA)},{\phi^F_{s_2}}^{(AA)}\right)$. We select location $TR2$, which is in the middle of the transition region, and has single-point intermittency about 0.4 (refer to figure \ref{Cf_sch}). 

\begin{figure}
    \centering
  \includegraphics[scale=0.39]{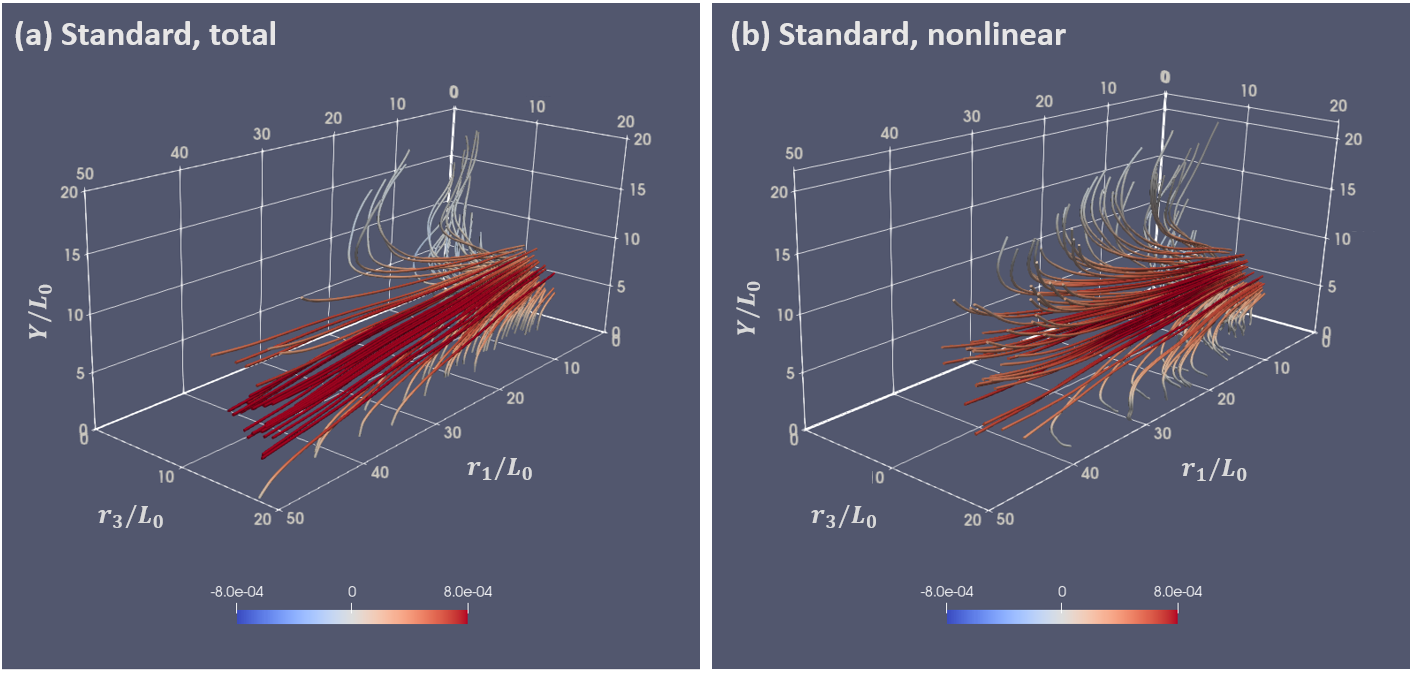}
  \includegraphics[scale=0.39]{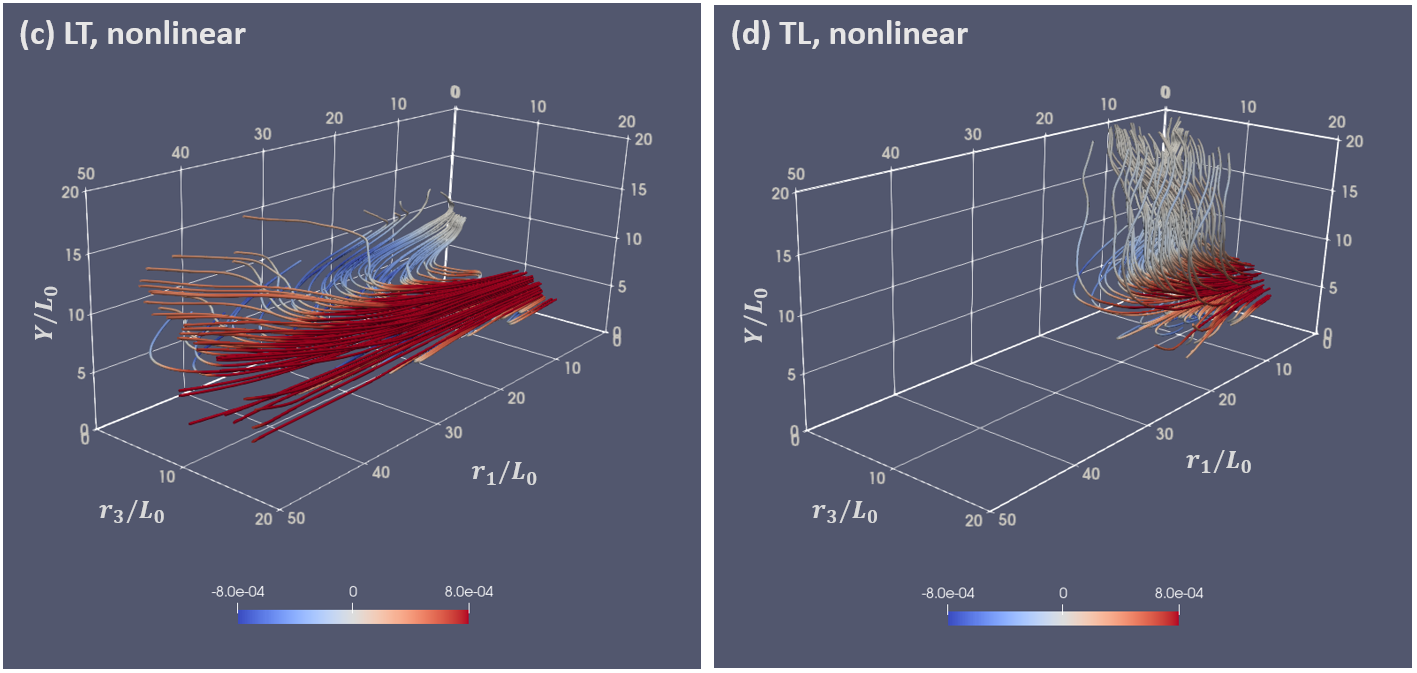}
  \includegraphics[scale=0.39]{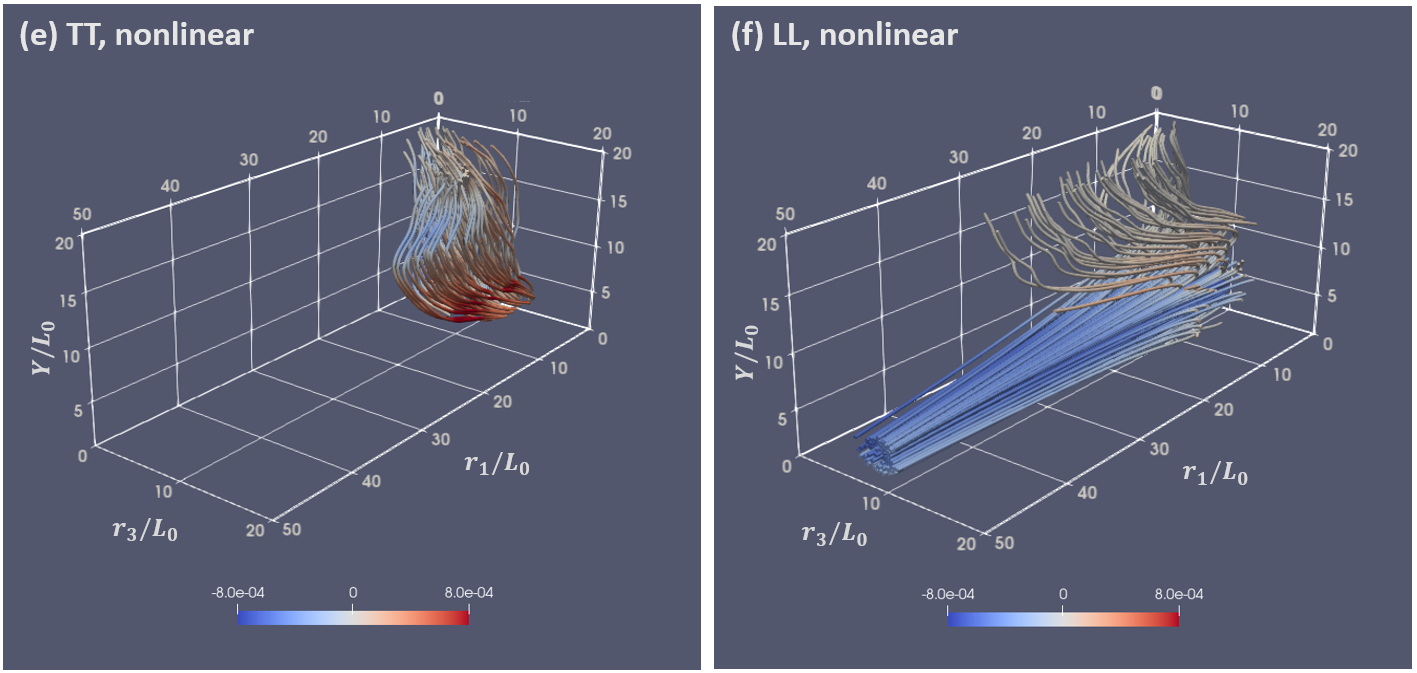}
    \caption{Stream-tubes of (a) the total flux vector  $\left({{\phi_{r_1}}},{{\phi_{r_3}}},{{\phi_{s_2}}}\right)$, (b) the standard-averaged non-linear flux vector $\left({\phi^F_{r_1}}, {\phi^F_{r_3}}, {\phi^F_{s_2}}\right)$, and the conditionally-averaged vectors (c)  $\left({{\phi^F_{r_1}}^{(LT)}}, {{\phi^F_{r_3}}^{(LT)}},{{\phi^F_{s_2}}^{(LT)}}\right)$, (d) $\left({{\phi^F_{r_1}}^{(TL)}}, {{\phi^F_{r_3}}^{(TL)}},{{\phi^F_{s_2}}^{(TL)}}\right)$, (e) $\left({{\phi^F_{r_1}}^{(TT)}}, {{\phi^F_{r_3}}^{(TT)}},{{\phi^F_{s_2}}^{(TT)}}\right)$ and (f) $\left({{\phi^F_{r_1}}^{(LL)}}, {{\phi^F_{r_3}}^{(LL)}},{{\phi^F_{s_2}}^{(LL)}}\right)$ at $TR2$. The plots are generated by placing a sphere of radius $5L_0$ at point $(r_1,r_3, Y)=(5L_0, 10L_0, 3L_0)$ and tracing the stream-tubes crossing the sphere. The stream-tubes are coloured according to the sign of the first component i.e.\ ${\phi_{r_1}}$, ${\phi^F_{r_1}}$,  ${\phi^F_{r_1}}^{(LT)}$, ${\phi^F_{r_1}}^{(TL)}$, ${\phi^F_{r_1}}^{(TT)}$ and ${\phi^F_{r_1}}^{(LL)}$  (red for positive, blue for negative, thus indicating inverse or forward cascade in the $r_1$ direction respectively). The colour bars also refer to the value of the first component (the min/max values are the same to facilitate comparison) 
    }
    \label{3D_con_Nonlinear}
\end{figure}

Stream-tubes obtained from the total and the non-linear flux vectors (standard or conditionally-averaged) are shown in figure \ref{3D_con_Nonlinear}; see caption for details. Panel (a) shows the standard-averaged total energy flux vector $\left(\phi_{r_1},\phi_{r_3},\phi_{s_2}\right)$ that includes the non-linear, linear, pressure, and viscous components, refer to equation \eqref{eq:flux_scale_space}. This plot is very similar to that of panel (b) that shows the non-linear component $\left(\phi^F_{r_1}, \phi^F_{r_3}, \phi^F_{s_2}\right)$, the latter being the dominant component (see \cite{yao2022analysis} for a detailed discussion on the other components). Both plots depict a dense cluster of stream-tubes with energy flowing to larger $r_1$ scales (inverse cascade) with energy originating at $Y \approx 5L_0$ and $r_3\approx 10L_0$. There is milder inverse cascade in larger $r_3$ scales. In another (smaller) cluster, energy flux vectors rotate and bend towards the Y-axis ($r_1=0,r_3=0$). 

The decomposition \eqref{final_phiF} allows us to probe in more detail the origin of the strong inverse cascade in the $r_1$ direction and identify the flow events that determine it. In the homogeneous spanwise direction $TL$ and $LT$ events can be combined together as $TL+LT$ (see figure \ref{sche_r3}). However, the streamwise direction is inhomogeneous and these events have to be considered separately. As mentioned earlier, for a fixed $(X,Y)$ spatial location and $r_3$ separation, a $TL$ event with $r_1>0$  takes place across the upstream laminar/turbulent interface of a spot. On the other hand, an $LT$ event is taken across the downstream interface, refer to figure \ref{sche_r1}.

Panels (c) and (d) of figure \ref{3D_con_Nonlinear} show the stream-tubes for the conditionally-averaged flux vectors $\left({\phi^F_{r_1}}^{(LT)}, {\phi^F_{r_3}}^{(LT)},{\phi^F_{s_2}}^{(LT)}\right)$  and $\left({\phi^F_{r_1}}^{(TL)}, {\phi^F_{r_3}}^{(TL)},{\phi^F_{s_2}}^{(TL)}\right)$ respectively. It is very clear that the $LT$ flux vector contributes most strongly to the inverse cascade; this corresponds to the downstream laminar/turbulent interface of a spot. Indeed there is a cluster of stream-tubes whose direction indicates the transfer of energy to larger streamwise scales, up to  $r_1 \approx 50L_0$. The stream-tubes then bend towards smaller scales, $r_1,r_3 \to 0$. On the other hand, $TL$ events that correspond to energy flux across the upstream interface of a turbulent spot, also contribute to inverse cascade, but over a  shorter streamwise range, $0<r_1<20L_0$. This clearly indicates that inter-scale energy transfer processes are different at the upstream and downstream interfaces of turbulent spots.

Panels (e) and (f) of figure \ref{3D_con_Nonlinear} show the stream-tubes of the conditionally-averaged flux vectors   $\left({\phi^F_{r_1}}^{(TT)}, {\phi^F_{r_3}}^{(TT)},{\phi^F_{s_2}}^{(TT)}\right)$ and $\left({\phi^F_{r_1}}^{(LL)}, {\phi^F_{r_3}}^{(LL)},{\phi^F_{s_2}}^{(LL)}\right)$. The former corresponds to interscale transfer within spots; there is weak inverse cascade and the stream-tubes bend towards smaller scales. The latter corresponds to time instants where both points are located within laminar regions.  There is strong forward cascade to small scales and then bending and energy transfer to larger scales away from the wall. This is a quite complicated energy flux pattern, which is difficult to interpret physically.

\begin{figure}
\centering
\includegraphics[scale=0.4]{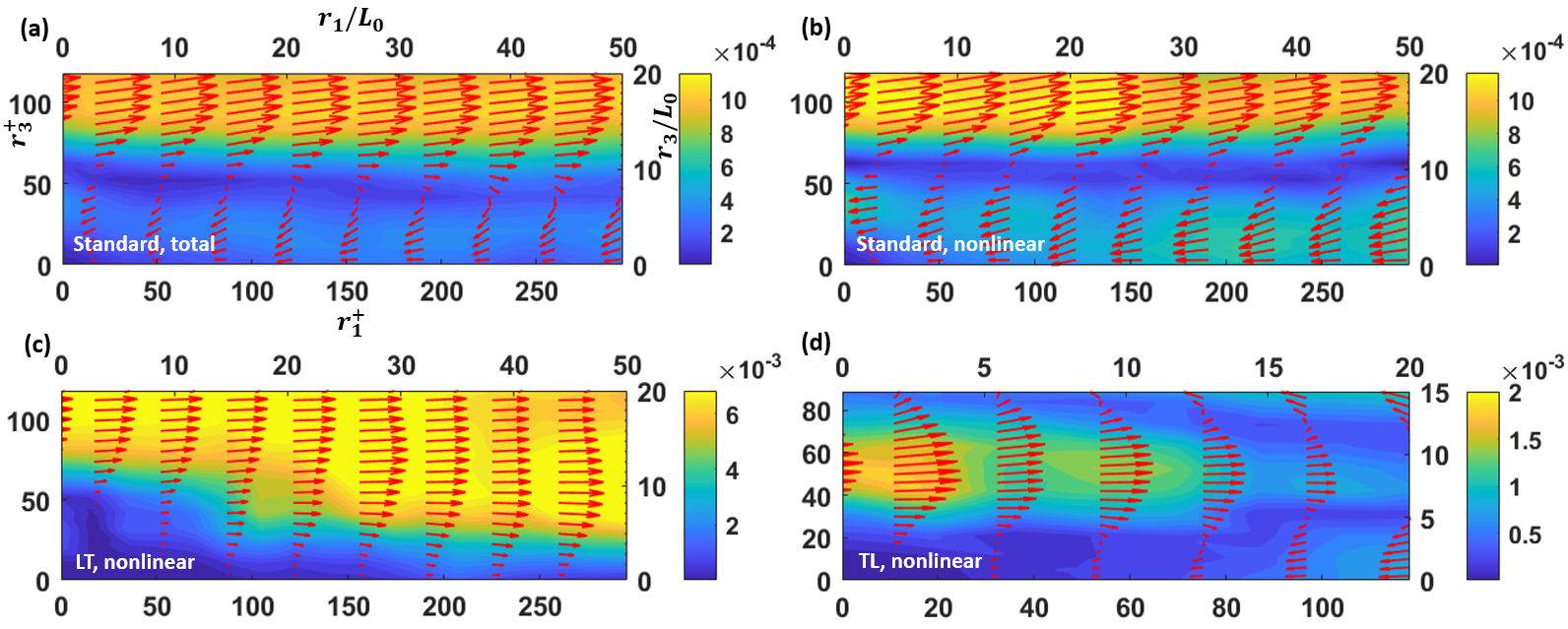}
\includegraphics[scale=0.41]{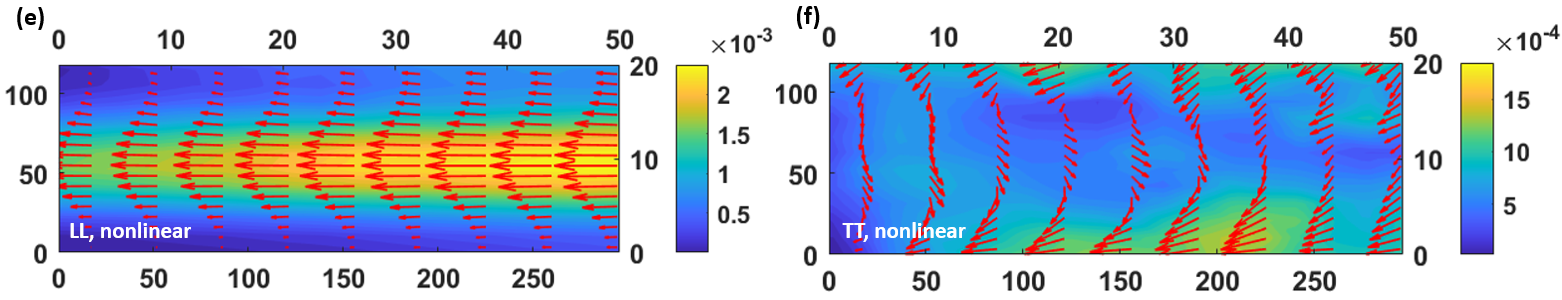}
    \caption{Flux vectors (a) $\left(\phi_{r_1},\phi_{r_3}\right)$,
    (b) $\left(\phi^F_{r_1}, \phi^F_{r_3}\right)$
    ,(c)  $\left({\phi^F_{r_1}}^{(LT)}, {\phi^F_{r_3}}^{(LT)}\right)$,
    (d)  $\left({\phi^F_{r_1}}^{(TL)}, {\phi^F_{r_3}}^{(TL)}\right)$,
    (e) $\left({\phi^F_{r_1}}^{(TT)}, {\phi^F_{r_3}}^{(TT)}\right)$ and
    (f)  $\left({\phi^F_{r_1}}^{(LL)}, {\phi^F_{r_3}}^{(LL)}\right)$ 
     on the $(r_1,r_3)$ plane at wall-normal height $Y=4.5L_0$ and location $TR2$. Contours represent the magnitude of the energy flux vectors in the plane.}
    \label{3D_con_Nonlinear_r1r3}
\end{figure}

Plots in the 3D hyperplane $(r_1,r_3, Y)$ visualise the main features of the energy flux paths, but can hide important detail. To uncover this detail, in figure  \ref{3D_con_Nonlinear_r1r3} we plot the flux vectors in the $(r_1,r_3)$ plane at the specific height $Y=4.5L_0$.  The total flux vector and the standard-averaged non-linear component (panels (a) and (b) respectively) are very similar and show a recirculating pattern with inverse cascade for $r_3>10$ over the range of $r_1$ examined and forward cascade for $r_3<10L_0$. It is very interesting to see that around $r_3=10L_0$ the energy flux is negligible; this cannot be easily observed from figure \ref{3D_con_Nonlinear}. The conditionally-averaged fluxes also show detail that cannot be discerned from the 3D plots. For example, the strong inverse cascade of ${\phi^F_{r_1}}^{(LT)}$ extends over the whole range of $r_3$ and $r_1$, while for  ${\phi^F_{r_1}}^{(TL)}$ it extends only in a specific range of separations, $r_3 \approx (5-15)L_0$, depending on $r_1$, as can be seen from panels (c) and (d) respectively. The flux component   ${\phi^F_{r_1}}^{(LL)}$ (panel (e)) clearly demonstrates forward cascade over the whole $r_3$ range examined, but ${\phi^F_{r_1}}^{(TT)}$ (panel (f)) is very small around $r_3\approx 10$ and increases at the boundaries of the domain. These plots confirm that the strong inverse cascade in the standard-averaged flux is due to $LT$ events at the downstream interface of the turbulent spots. Interestingly, the energy fluxes due to $TL$ and $LL$ events almost cancel out around $r_3\approx 10L_0$, and this explains the very small fluxes in this area for  ${\phi^F_{r_1}}$. Notice also that the $TT$ events account for the forward cascade observed in ${\phi^F_{r_1}}$ and ${\phi_{r_1}}$ for small $r_3$ separations.      

\begin{figure}
    \centering
 \includegraphics[scale=0.45]{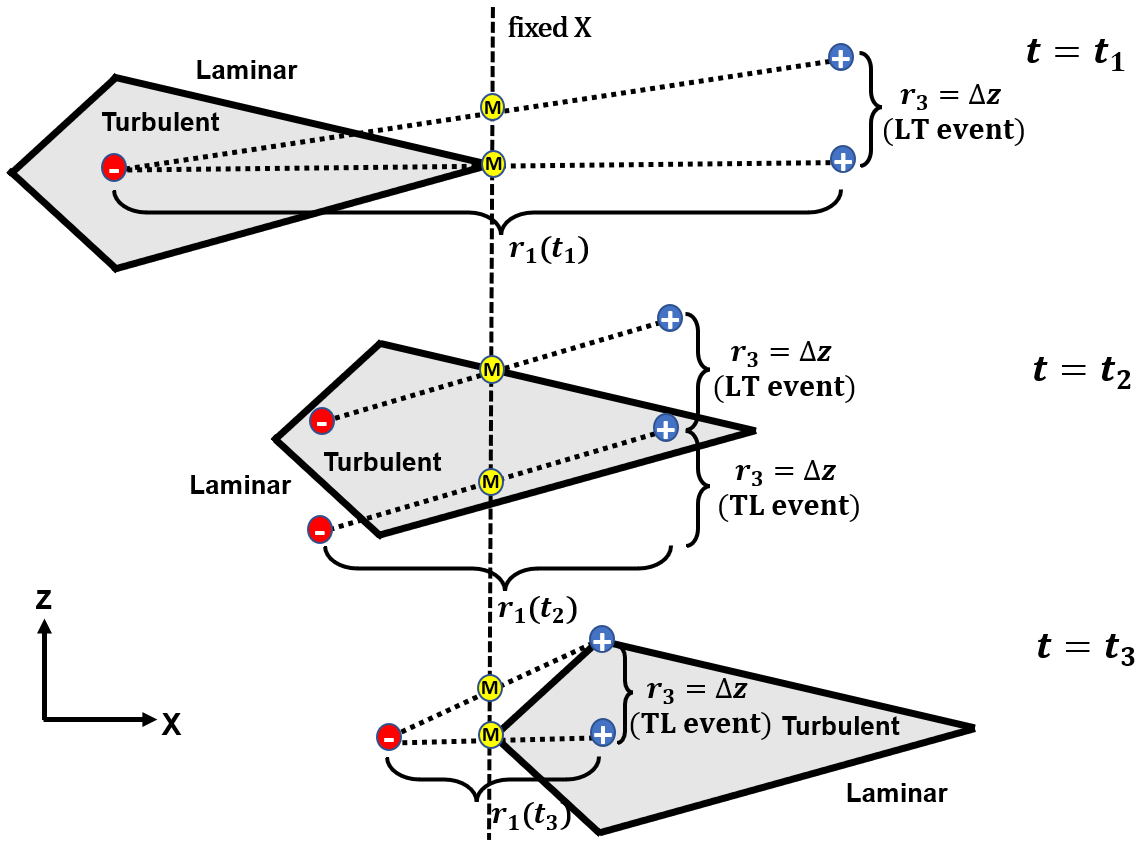}
    \caption{A propagating turbulent spot of diamond shape crossing a fixed $X$ location; the view shown is in the $(X, Z)$ plane. Red, yellow and blue dots represent the $x_i^-$, $X_i$, and $x_i^+$ points respectively. Only points with fixed $r_3=\Delta z$ separation are shown.}
    \label{TL_SCH_ver2}
\end{figure}

\begin{figure}
    \centering
  \includegraphics[scale=0.4]{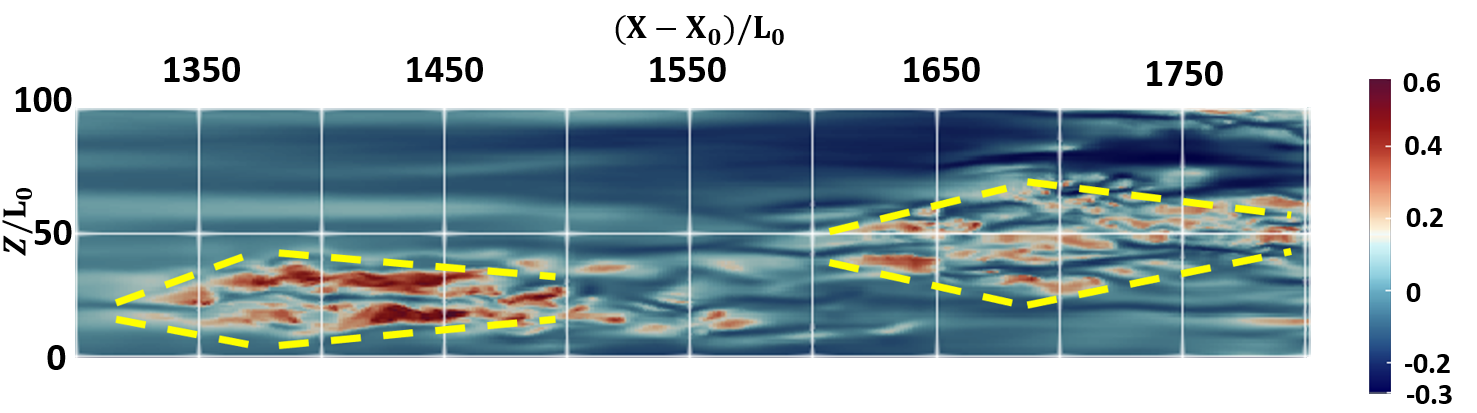}
    \caption{Contour plot of instantaneous streamwise velocity fluctuations. The boundaries of two turbulent spots in the transition region (marked by dashed yellow lines) indicate spots of approximately diamond shape.}
    \label{Sch_r13increase}
\end{figure}

The above figures have demonstrated the central role of the downstream laminar/turbulent interface on the interscale transfer and in particular the inverse cascade over a large range of scales. Additionally, $TL$ events were localised in a smaller range of spanwise and streamwise separations. We now try to explain physically this behaviour with the aid of the cartoons shown in figure \ref{TL_SCH_ver2}. More specifically, we consider a fixed $X$ location (for figures \ref{3D_con_Nonlinear} and \ref{3D_con_Nonlinear_r1r3},  $X=X_{TR2}$) and follow a turbulent spot of diamond shape as it propagates to the right and crosses this location. The arrowhead shape at the upstream and downstream ends is a simplified, but rather realistic, approximation. This can be seen from figure \ref{Sch_r13increase} where we demarcate the boundaries of two turbulent spots. It is also consistent with experimental spot observations, refer to figures 12 and 14 in \cite{anthony2005high}. The sharp corners of the spot around the maximum thickness are less realistic; the shape is more rounded in this region as can be seen from figure 18 of \cite{marxen2019turbulence}. However, analysis of this simplified shape can provide significant physical insight, as will be seen next.

In figure \ref{TL_SCH_ver2}, snapshots of the  propagating spot at three time instants $t_1, t_2$ and $t_3$ are shown. In all snapshots, we consider a fixed middle point (denoted with a yellow dot) located at the streamwise position, $X$. The blue and red dots represent the $x_i^+$ and $x_i^-$ points respectively. Only points with a fixed spanwise separation, $r_3=\Delta z$, are shown in the figure. At the time $t=t_1$, the downstream apex of the spot lies exactly at the fixed $X$ location. It can be seen that for an $LT$ event (the only type of event possible at this time instant), the streamwise separation $r_1(t_1)$ is very long, of the order of the spot length. For $\Delta z=0$, $r_1(t_1)$ attains a maximum value, equal to twice the spot length.

At time instant $t=t_2$, approximately half of the spot has crossed $X$. The valid spanwise locations of the middle point are determined by the spreading angle of the front apex. Note that for fixed $r_3=\Delta z$, $TL$ and $LT$ events co-exist, but it is clear that the $r_1(t_2)$ separation of an $LT$ event is shorter to the one at $t=t_1$, i.e.\ $r_1(t_2)<r_1(t_1)$. At $t=t_3$, the whole spot has crossed the considered location, thus the rear apex is at $X$. At this time instant, only $TL$ events are possible. It can be seen that only a narrow range of $r_1$ separations is admissible for a given $r_3$. The actual range depends on the spreading angle of the rear apex. Thus on average, the valid $r_1$ separations corresponding to $TL$ events at $t_2$ and $t_3$ is more narrow compared to $LT$ events at $t=t_1$ and $t_2$. This explains the inverse cascade over a wider range of separations for $LT$ events shown in panel \ref {3D_con_Nonlinear_r1r3}(c). We also conjecture that the largest admissible $r_1$ value mentioned earlier explains why the stream-tubes shown in figure \ref{3D_con_Nonlinear}(c) reach up to a maximum $r_1$ (the exact value depends on point of origin of the stream-tubes in the $(r_3, Y)$ plane) and then bend backward towards small scales. 

\begin{figure}
    \centering
  \includegraphics[scale=0.4]{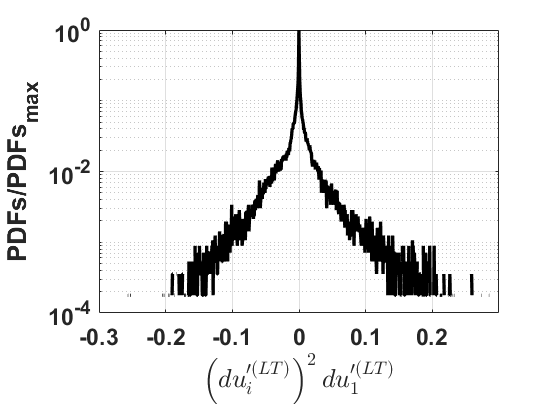}
  \includegraphics[scale=0.4]{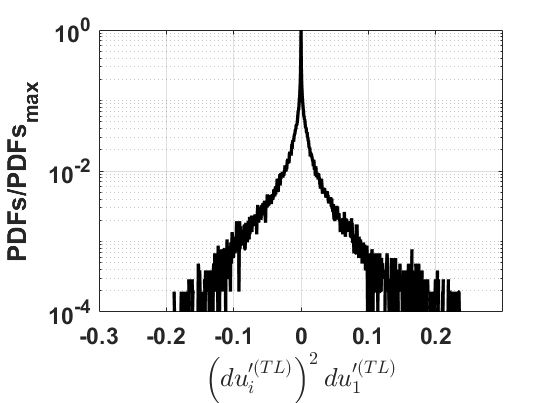}
  \includegraphics[scale=0.4]{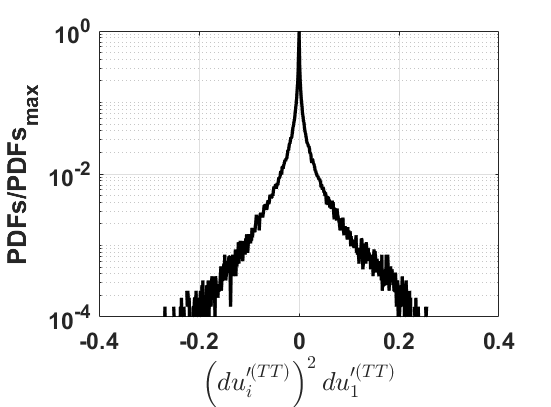}
  \includegraphics[scale=0.4]{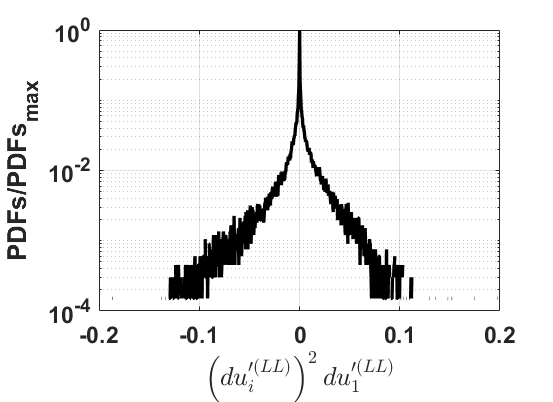}
    \caption{PDFs of $\left(du'^{(AA)}_i\right)^2du'^{(AA)}_1$ for $AA=$ $LT$, $TL$, $TT$ and $LL$ for $r_1=30L_0$ and $r_3=10L_0$ at $Y=4.5L_0$.}
    \label{fig:pdf_cond_r1_fluxes}
\end{figure}

To provide further insight into the observed behaviour of the conditionally-averaged fluxes, we examine in more detail the flux vector component in the $r_1$ direction ${\phi^F_{r_1}}^{(AA)}=\overline{\left(du'^{(AA)}_i\right)^2du'^{(AA)}_1}^{(AA)}$. In figure  \ref{fig:pdf_cond_r1_fluxes}, the probability density functions (PDFs) of  $\left(du'^{(AA)}_i\right)^2du'^{(AA)}_1$ for $AA=$ $LT$, $TL$,  $TT$ and $LL$ are plotted for separations $r_1=30L_0$ and $r_3=10L_0$ at plane $Y=4.5L_0$. The PDF of the instantaneous flux  $\left(du'^{(LT)}_i\right)^2du'^{(LT)}_1$ is asymmetric and skewed to positive values (implying inverse cascade after time-averaging). This is also the case but it is less evident for $\left(du'^{(TL)}_i\right)^2du'^{(TL)}_1$, while $\left(du'^{(TT)}_i\right)^2du'^{(TT)}_1$ is almost symmetric, and $\left(du'^{(LL)}_i\right)^2du'^{(LL)}_1$ is skewed to the left. Note the large positive and negative fluctuations of the instantaneous fluxes compared to the time-average values reported in figure \ref{3D_con_Nonlinear_r1r3}. This means that instantaneously energy flows in either direction, and intense fluxes of relatively low probability tip the balance in one direction or another after time-averaging.

\section{Conditionally-averaged scale energy production and transfer within a turbulent spot and comparison with fully developed turbulence}\label{sec:conditional_averaged_production}

The conditional averaging operations defined in section \ref{sec:conditional_operations} allow us to compute the scale energy production and interscale transfer within a turbulent spot and compare them with the corresponding quantities in the fully turbulent region. Similar work has been done for single-point statistics, for example, the turbulent kinetic energy by \cite{park2012boundary,Nolan_Zaki_2013,marxen2019turbulence}. To the best of our knowledge, this is the first time this type of analysis is extended to two-point statistics. 

We start by deriving a decomposition similar to \eqref{final_phiF} for the production term $\mathcal{P}$ of the KHMH equation \eqref{KHMH_standard}. This term consists of two components, one due to inhomogeneity of the mean flow in scale space, $\mathcal{P}_r=-2\overline{du'_i du'_{j}}\frac{\partial dU_i}{\partial r_j}$, and the other due to inhomogeneity in physical space, $\mathcal{P}_s=-2\overline{ du'_i u'^*_{j}} \frac{\partial dU_i}{\partial X_j}$. Here we decompose the former component, $\mathcal{P}_r$, which is the dominant one; this is also discussed below. Applying a process similar to that described in section \ref{sec:cond_decompostition}, we obtain

\begin{equation}
\begin{aligned}
&\mathcal{P}_r=-2\overline{du'_i du'_{j}}\frac{\partial dU_i}{\partial r_j}= \\
& -2\left (\gamma^{(TT)}\right)^2\overline{du'^{(TT)}_idu'^{(TT)}_{j}}^{(TT)}\frac{\partial dU_i^{(TT)}}{\partial r_j}
-
2\left ( \gamma^{(TL)}\right)^2\overline{du'^{(TL)}_idu'^{(TL)}_{j}}^{(TL)}\frac{\partial dU_i^{(TL)}}{\partial r_j}
\\
&-2\left(\gamma^{(LT)}\right)^2\overline{du'^{(LT)}_idu'^{(LT)}_{j}}^{(LT)}\frac{\partial dU_i^{(LT)}}{\partial r_j}
-
2\left (\gamma^{(LL)}\right)^2 \overline{du'^{(LL)}_idu'^{(LL)}_{j}}^{(LL)}\frac{\partial
dU_i^{(LL)}}{\partial r_j} \\
&- \phi,
\end{aligned}
\label{Pro_condi}
\end{equation}
where the additional term $\phi$ is given by
\begin{equation}
\begin{aligned}
&\phi= \\
&2\left (\gamma^{(TT)}\right )^2 dU_i^{(TT)}dU_{j}^{(TT)}\frac{\partial dU_i^{(TT)}}{\partial r_j}
+
2\left (\gamma^{(TL)}\right )^2 dU_i^{(TL)}dU_{j}^{(TL)}\frac{\partial dU_i^{(TL)}}{\partial r_j}\\
&+
2\left (\gamma^{(LT)}\right )^2 dU_i^{(LT)}dU_{j}^{(LT)}\frac{\partial dU_i^{(LT)}}{\partial r_j}
+
2\left(\gamma^{(LL)}\right)^2 dU_i^{(LL)} dU_{j}^{(LL)}\frac{\partial dU_i^{(LL)}}{\partial r_j}
\\
&+2\gamma^{(TT)}\overline{du_idu_{j}}^{(TT)}\left[ dU_i^{(TT)}\frac{\partial \gamma^{(TT)}}{\partial r_j}+\frac{\partial}{\partial r_j}\left( \gamma^{(LL)}dU_i^{(LL)}
+\gamma^{(TL)}dU_i^{(TL)}
+\gamma^{(LT)}dU_i^{(LT)}
\right)\right]
\\
&+2\gamma^{(TL)}\overline{du_idu_{j}}^{(TL)}\left[dU_i^{(TL)}\frac{\partial \gamma^{(TL)}}{\partial r_j}+\frac{\partial}{\partial r_j}\left( \gamma^{(LL)}dU_i^{(LL)}
+\gamma^{(TT)}dU_i^{(TT)}
+\gamma^{(LT)}dU_i^{(LT)}
\right)\right]\\
&+2\gamma^{(LT)}\overline{du_idu_{j}}^{(LT)}\left[dU_i^{(LT)}\frac{\partial \gamma^{(LT)}}{\partial r_j}+\frac{\partial}{\partial r_j}\left( \gamma^{(LL)}dU_i^{(LL)}
+\gamma^{(TL)}dU_i^{(TL)}
+\gamma^{(TT)}dU_i^{(TT)}
\right)\right]\\
&+2\gamma^{(LL)}\overline{du_idu_{j}}^{(LL)}\left[dU_i^{(LL)}\frac{\partial \gamma^{(LL)}}{\partial r_j}+\frac{\partial}{\partial r_j}\left( \gamma^{(TT)}dU_i^{(TT)}
+\gamma^{(TL)}dU_i^{(TL)}
+\gamma^{(LT)}dU_i^{(LT)}
\right)\right]\\
&-2dU_idU_{j}\frac{\partial dU_i}{\partial r_j}
\end{aligned}
\end{equation}

Note that the terms $\mathcal{P}_r^{(AA)}=-2\overline{du'^{(AA)}_idu'^{(AA)}_{j}}^{(AA)}\frac{\partial dU_i^{(AA)}}{\partial r_j}$ that appear in \eqref{Pro_condi} are the production terms of the conditionally-averaged KHMH equation \eqref{KH_CON}. Due to the strong shear in the wall-normal direction, the dominant component of $\mathcal{P}_r=-2\overline{du'_i du'_{j}}\frac{\partial dU_i}{\partial r_j}$ is $\mathcal{P}_{r(1,2)}=-2\overline{du'_1 du'_{2}} \left. \frac{\partial dU_1}{\partial r_2} \right |_{r_2=0}$. It can be easily proved (see \cite{yao2022analysis}) that for $r_1=0$ this component is equal to $-2\overline{du'_1 du'_{2}} \frac{\partial U_1}{\partial x_2}$. The corresponding production component due to  inhomogeneity of the mean flow in physical space, $\mathcal{P}_s=-2\overline{ du'_1 u'^*_{2}} \frac{\partial dU_1}{\partial X_2}$, is much smaller. Therefore in this section we consider only $\mathcal{P}_{r(1,2)}=-2\overline{du'_1 du'_{2}} \left. \frac{\partial dU_1}{\partial r_2} \right |_{r_2=0}$ and the corresponding conditionally-averaged $(TT)$ component $\mathcal{P}_{r(1,2)}^{(TT)}=-2\overline{du'^{(TT)}_1du'^{(TT)}_{2}}^{(TT)} \left. \frac{\partial dU_1^{(TT)}}{\partial r_2} \right |_{r_2=0}$ within a turbulent spot. We focus at location $TR2$ and compare the aforementioned component with  $\mathcal{P}_{r(1,2)}$ evaluated at the fully turbulent region, $TU$.

We also extend the decomposition \eqref{final_phiF} to the  total fluxes in physical and scales spaces $\boldsymbol{\phi_s}$ and  $\boldsymbol{\phi_r}$, defined in equations \eqref{eq:flux_phys_space} and \eqref{eq:flux_scale_space} respectively. The resulting  expressions are 
\begin{equation}
\boldsymbol{\phi_s}_j= \gamma^{(TT)}\boldsymbol{\phi_s}^{(TT)}_j +
\gamma^{(TL)}\boldsymbol{\phi_s}^{(TL)}_j+
\gamma^{(LT)}\boldsymbol{\phi_s}^{(LT)}_j+
\gamma^{(LL)}\boldsymbol{\phi_s}^{(LL)}_j+
\phi,
\label{eq:decomposition_phi_s}
\end{equation}
where
\begin{equation}
\begin{aligned}
&\boldsymbol{\phi_s}_j^{(AA)}=\overline{{U^*_j}^{(AA)}d q^{2^{(AA)}}}^{(AA)}
+
\overline{{u'^*_j}^{(AA)}d q^{2^{(AA)}}}^{(AA)} 
+
\overline{2 d u'^{(AA)}_j d p'^{(AA)} }^{(AA)} 
-
\frac{1}{2} \nu  \overline{\frac{\partial  d q^{2^{(AA)}} }{\partial X_j}}^{(AA)},
\end{aligned}
\label{eq:phi_s_AA}
\end{equation} 

and

\begin{equation}
\boldsymbol{\phi_r}_j= \gamma^{(TT)}\boldsymbol{\phi_r}^{(TT)}_j+
\gamma^{(TL)}\boldsymbol{\phi_r}^{(TL)}_j+
\gamma^{(LT)}\boldsymbol{\phi_r}^{(LT)}_j+
\gamma^{(LL)}\boldsymbol{\phi_r}^{(LL)}_j+
\phi,
\label{eq:decomposition_phi_r}
\end{equation}
where
\begin{equation}
\boldsymbol{\phi_r}_j^{(AA)}=
\overline{ d u'^{(AA)}_j d q^{2^{(AA)}} }^{(AA)}
+
\overline{ dU_j^{(AA)} d q^{2^{(AA)}} }^{(AA)}
-
2\nu \overline{\frac{\partial d q^{2^{(AA)}} }{\partial r_j}}^{(AA)}.
\label{eq:phi_r_AA}
\end{equation}
The full expressions, including the remainder terms, are provided in Appendix \ref{sec:appendix_total_fluxes}.  

\subsection{Conditionally- and standard-averaged production and fluxes}

\begin{figure}
    \centering
    \includegraphics[scale=0.42]{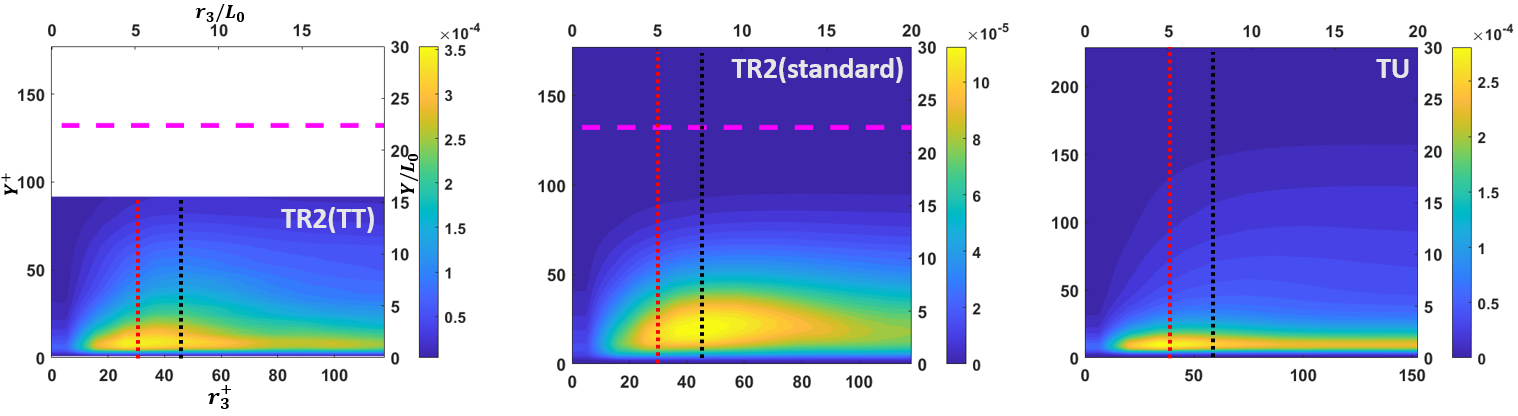}  
    \caption{Contour plots of conditionally-averaged production  $\mathcal{P}_{r(1,2)}^{(TT)}$ at $TR2$ (left), standard-averaged production $\mathcal{P}_{r(1,2)}$ at $TR2$ (middle) and  $TU$ (right) on the $(r_3,Y)$ plane for $r_1=0$. The red and black vertical dotted lines are placed at $r_3=5L_0$ and $7.5L_0$ respectively. The horizontal purple line indicates the local boundary layer thickness.}
    \label{2D_pro}
\end{figure}

\begin{figure}
    \centering
    \includegraphics[scale=0.4]{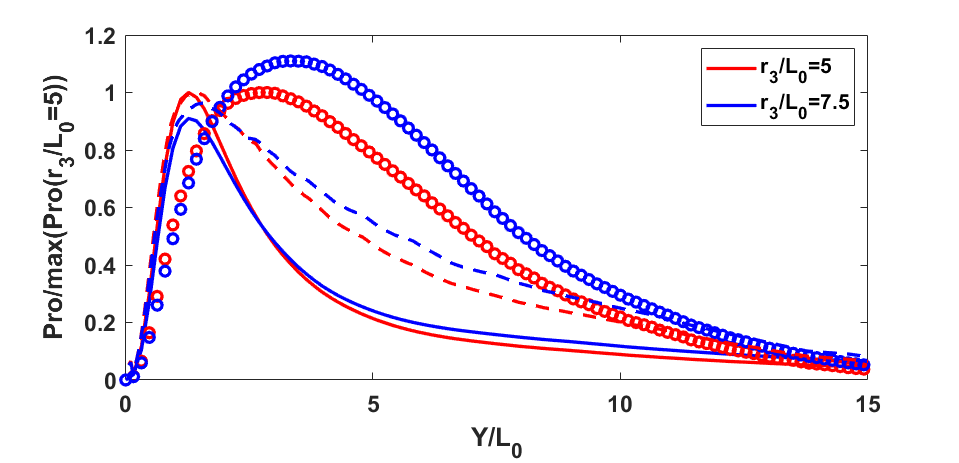}
    \caption{Variation of production terms along the wall-normal distance at $r_3/L_0=5.0$ and $7.5$. The variation is along the dotted vertical lines shown in figure \ref{2D_pro} that pass through the corresponding production peaks. The plots are normalised by the value at $r_3/L_0=5.0$. The solid lines represent  $\mathcal{P}_{r(1,2)}$ at $TU$. The dashed lines denote  $\mathcal{P}_{r(1,2)}^{(TT)}$ and the circles $\mathcal{P}_{r(1,2)}$, both at location $TR2$.}
    \label{HandVlines}
\end{figure}

In figure \ref{2D_pro}, contours of  $\mathcal{P}_{r(1,2)}^{(TT)}$ at $TR2$ and of $\mathcal{P}_{r(1,2)}$ at $TR2$ and $TU$ are plotted in the $(r_3,Y)$ plane for $r_1=0$. It can be seen that the production peaks within the turbulent spot and the fully turbulent region are located at approximately the same spanwise separation and wall-normal height, $r_3\approx5L_0$ and $Y \approx 1.3L_0$ respectively. On the other hand, the peak of  $\mathcal{P}_{r(1,2)}$ at $TR2$ is found to be at larger $r_3$ separation and further away from the wall,  $r_3 \approx 7.5L_0$ and $Y \approx 4.5L_0$. We mark the spanwise scales where the peaks appear, $r_3=5L_0$ and $7.5L_0$, with vertical dotted lines in figure \ref{2D_pro}, and plot the variation of the three production terms along these lines in figure \ref{HandVlines}. Notice the very close matching of the conditionally-average production $\mathcal{P}_{r(1,2)}^{(TT)}$ at $TR2$ (dashed line) and the standard-averaged production $\mathcal{P}_{r(1,2)}$ at $TU$ (solid line) close to the wall (for $Y\le 1.3L_0$) while further away the two sets deviate.  On the other hand, $\mathcal{P}_{r(1,2)}$ at $TR2$ shows significantly different behaviour even close to the wall, and of course peaks at a different distance. 

\begin{figure}
    \centering
\includegraphics[scale=0.4]{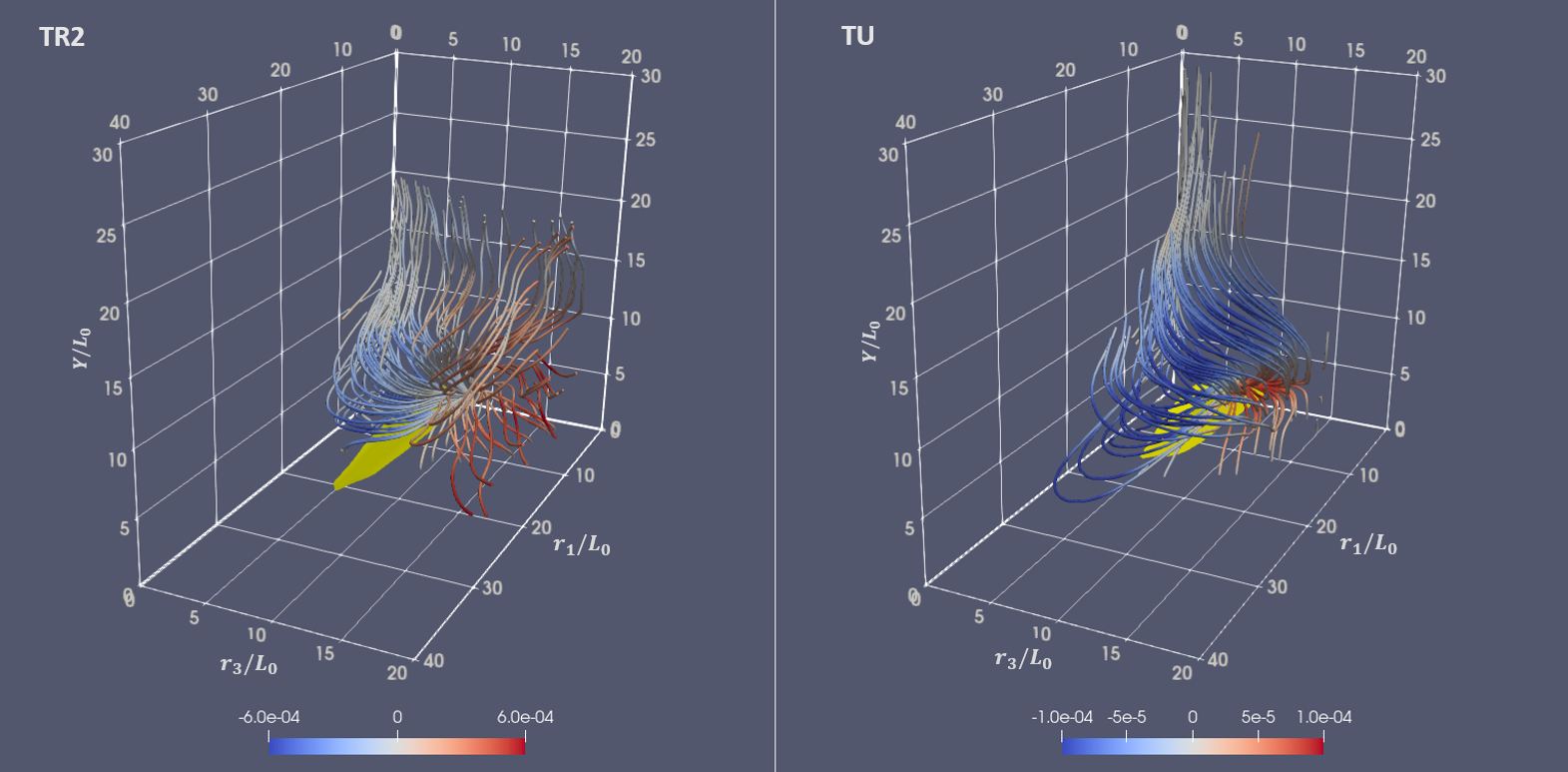}
    \caption{Stream-tubes of the conditionally-averaged total flux vector $\left({\phi_{r_1}^{(TT)}}, {\phi_{r_3}^{(TT)}},{\phi_{s_2}^{(TT)}}\right)$ at $TR2$ (left) and of the standard-averaged vector $\left({{\phi_{r_1}}},{{\phi_{r_3}}},{{\phi_{s_2}}}\right)$ at $TU$ in the 3D ($r_1,r_3,Y$) hyper-plane. The stream-tubes are colored according to the sign of $\phi_{r_3}$ (red for positive, blue for negative, thus indicating inverse and forward cascade respectively). The color bar refers to the value of $\phi_{r_3}$. The plots were generated by placing a sphere of radius $15L_0$ at  $(r_1=5L_0, r_3=5L_0, Y=1.3L_0)$ and tracing the stream-tubes crossing the sphere. Isosurfaces of the production term with values $0.9\times\max \left(\mathcal{P}_{r(1,2)}^{(TT)}\right)$(left) and $0.85\times\max \left(\mathcal{P}_{r(1,2)}\right)$(right) are shown in yellow.
    }    
    \label{3Dflux}
\end{figure}

\begin{figure}
    \centering
    \includegraphics[scale=0.4]{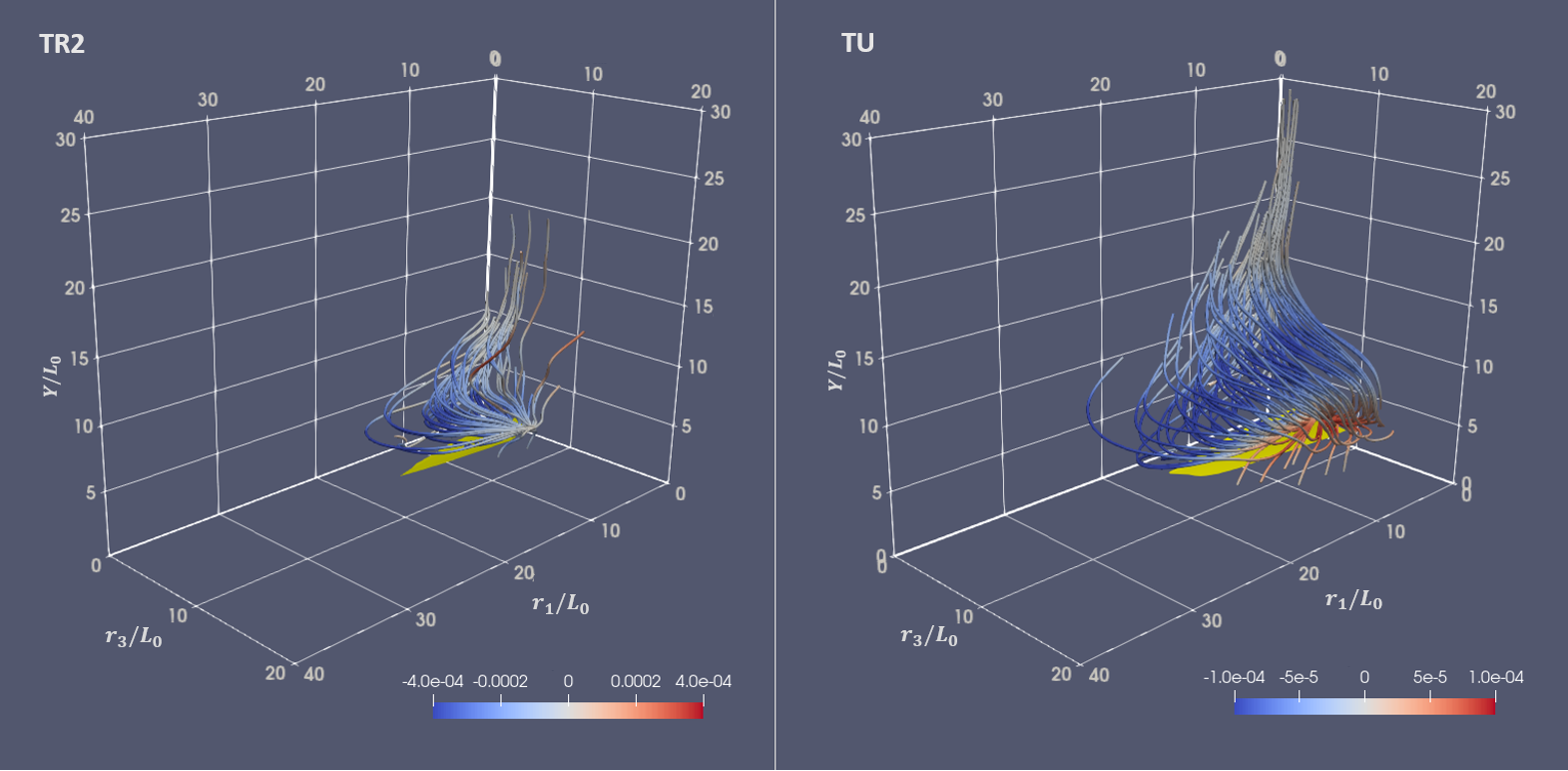}
    \caption{Same as figure \ref{3Dflux}, but the sphere is now placed at  $(r_1=5L_0, r_3=2.5L_0, Y=1.3L_0)$ and has smaller radius, $4L_0$. Isosurfaces of the production term with values  $0.95\times \max \left(\mathcal{P}_{r(1,2)}^{(TT)}\right)$(left) and $0.85\times \max \left(\mathcal{P}_{r(1,2)}\right)$(right) are shown in yellow.}
    \label{3Dflux_small}
\end{figure}

The left panel of figure \ref{3Dflux} shows stream-tubes in the $(r_1,r_3,Y)$ hyperplane obtained from the conditionally-averaged total fluxes $\left({\phi_{r_1}^{(TT)}}, {\phi_{r_3}^{(TT)}},{\phi_{s_2}^{(TT)}}\right)$ at $TR2$ together with an isosurface of the conditionally-averaged production.  On the right panel we plot $\left({{\phi_{r_1}}},{{\phi_{r_3}}},{{\phi_{s_2}}}\right)$ and production in the fully turbulent region. The latter figure reflects the dynamics of near-wall turbulence; energy is extracted from the mean flow at the buffer layer where the production peak is located, then it is transferred away from the wall and towards larger $r_1$ scales before bending back to smaller scales (dissipation region). This behaviour is related to the self-sustained turbulence mechanism near the wall, see  \cite{cimarelli2013paths}. A similar pattern can be discerned in the left panel, but the inverse cascade and flow of energy away from the wall is over a smaller range of $r_1$ separations (up to $r_1\approx 10$); the stream-tubes again bend towards small scales. There are also some deviations between the two plots for larger $r_3$ separations. If the centre of the sphere (used for identifying which stream-tubes to trace) is placed at smaller $r_3$ and the radius is reduced, the similarity between the two panels is more evident, see figure \ref{3Dflux_small} and caption for details.

The shorter range of inverse cascade in the $TR2$ location compared to the $TU$ is probably because the spots are still developing, the merging is not yet complete, thus they have a smaller footprint in the streamwise direction.  Note also the similarities of the left panels of figures \ref{3Dflux} and \ref{3Dflux_small} with the bottom left panel of figure \ref{3D_con_Nonlinear} that shows only the non-linear component of the flux vector. This similarity confirms that this is the most important component that determines the overall behaviour.  

\section{Conclusions}\label{sec:conclusions}
We apply conditional averaging to study the interscale energy transfer process during bypass transition. To this end, we define two-point intermittencies and apply them to decompose the energy fluxes into different components that depend on the local conditions at the two points used to define the flux; the points are both within a laminar region or a turbulent spot or straddle the laminar/turbulent interface. The flux terms are evaluated numerically directly in the scale space because conditional averaging does not commute with the spatial derivative operator.


In the $(r_1, r_3, Y)$ hyper-plane, strong inverse cascade is found in the $r_1$ direction, due to the non-linear fluxes across the downstream and upstream boundaries of a spot (head and tail respectively). For the former boundary, the inverse cascade extends over a larger range of $r_1$ separations compared to the latter boundary. We explain this finding by considering a propagating spot as it passes across a fixed streamwise location.  

We derive also the conditionally-averaged KHMH equation and consider the production term and the total energy fluxes when both points are located within a turbulent spot ($TT$ events). We compare with the corresponding terms in the fully turbulent region and find significant similarities, but also some differences. In both plots, a cluster of stream-tubes originates from the production peak and transfers energy to larger scales before bending back to small scales and the near-wall region. This spiral shape is similar to that found in the fully turbulent region and in channel flow. However, the extent of the spiralling motion is confined to smaller separations, probably because the spots have not fully merged yet. Also, a smaller cluster of stream-tubes transfers energy in the $r_1$ and $r_3$ directions, which is not found in the fully turbulent region. 

The conditional averaging approach for two-point statistics developed in the paper can be applied to other flow configurations that exhibit sharp interfaces, such as wakes and jets, where a turbulent/non-turbulent interface separates the irrotational and vortical regions. Important questions remain to be answered, for example, do the conditionally-averaged statistics exhibit self-similarity? How does this develop as the jet/wake expands? Research in this direction is left as future work.  

\section*{Acknowledgements}
H. Yao acknowledges financial support from the Dept. of Aeronautics, Imperial College London and the Imperial College-CSC scholarship. The authors also wish to acknowledge the UK Turbulence Consortium (UKTC) for providing access to the ARCHER high performance computing facility through EPSRC grant EP/R029326/1.

\section*{Declaration of Interests.} The authors report no conflict of interest.

\appendix
\section{Calculation of derivatives of two-point quantities directly in scale space}\label{sec:appendix_numerical_calculation}
In this Appendix, we provide the steps for the numerical evaluation of derivatives of two-point quantities directly in scale space.

\subsection{Calculation of $\left. \frac{\partial  dU_i^{(AA)}}{\partial r_2} \right |_{r_2=0} $}

\begin{figure}
    \centering
  \includegraphics[scale=0.4]{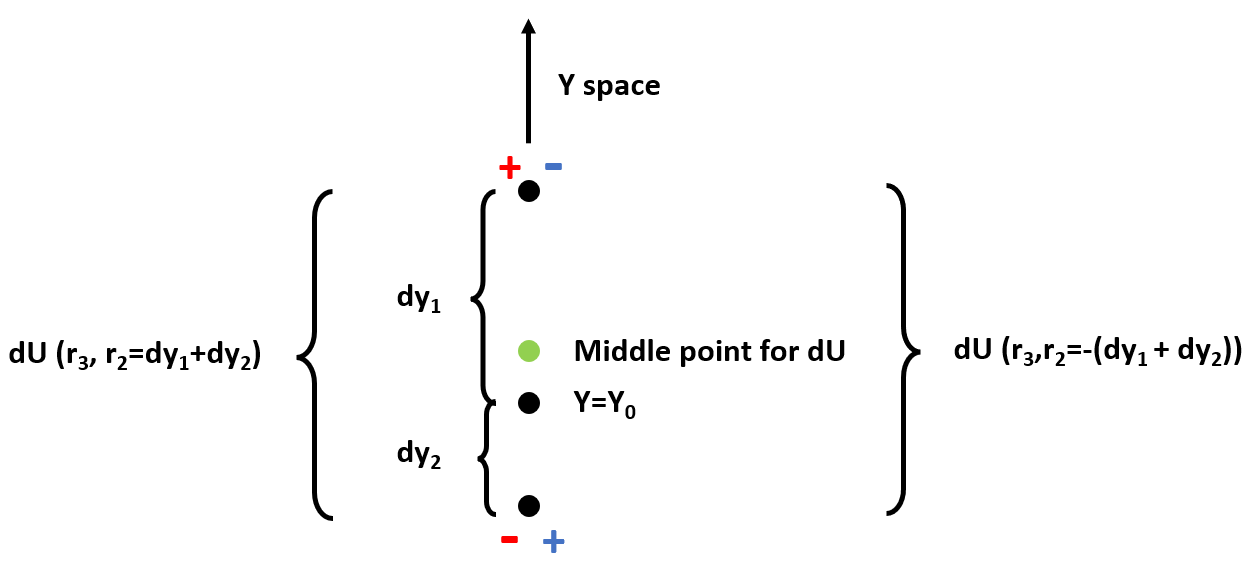}
    \caption{Sketch for the calculation of  $\left. \frac{\partial  dU_i^{(AA)}}{\partial r_2} \right |_{r_2=0}$.}
    \label{dudr2}
\end{figure}

\begin{enumerate}
\item Consider a cell with centroid at distance  $Y=Y_0$ from the wall. The distances of the centroids of the cells located above and below are $dy_1$ and  $dy_2$ respectively, refer to figure \ref{dudr2}. Recall that the grid is non-uniform in the wall-normal direction, so $dy_1 \ne dy_2$. Calculate $dU_i^{(AA)}(r_2=dy_1+dy_2)=U_i^{(AA)}\left(Y_0+dy_1\right)-U_i^{(AA)}\left(Y_0-dy_2\right)$ i.e.\ the velocity difference between the red $+/-$ markers in figure \ref{dudr2}.
\item Calculate $dU_i^{(AA)}\left(r_2=-(dy_1+dy_2)\right)=-dU_i^{(AA)}(r_2=dy_1+dy_2)$ i.e.\ the velocity difference between the blue $+/-$ markers.
\item Use central difference scheme to compute
\begin{equation}
 \left. \frac{\partial  dU_i^{(AA)}}{\partial r_2} \right |_{r_2=0} =
\frac{ dU_i^{(AA)}(r_2=dy_1+dy_2) -dU_i^{(AA)}(r_2=-(dy_1+dy_2))}{2(dy_1+dy_2)}
\end{equation}
and store the value at the midpoint (denoted with a green dot in figure \ref{dudr2}).
\item Repeat steps (i)-(iii) for cell centroids at different heights $Y_0$.
\item Because the mesh is non-uniform, the middle point is not located at $Y_0$, so interpolate values of $\frac{\partial  dU_i^{(AA)}}{\partial r_2}|_{r_2=0}$ at midpoints to obtain value at $Y_0$.
\end{enumerate}

\begin{figure}
    \centering
\includegraphics[scale=0.4]{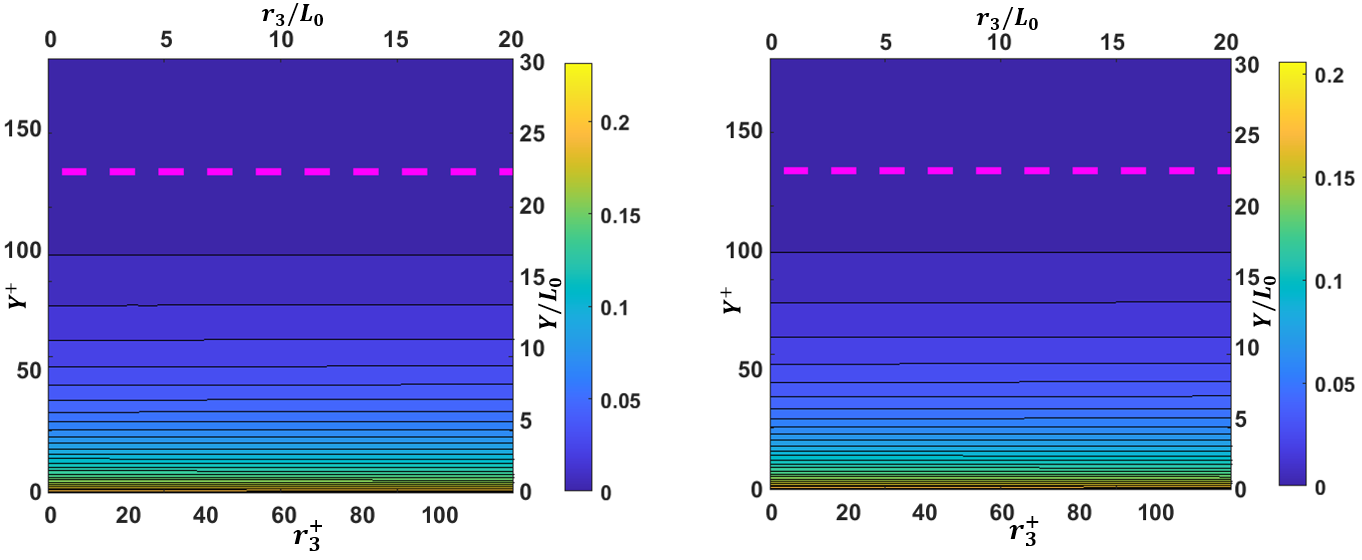} 
    \caption{Contour plot of $\left. \frac{\partial  dU_1}{\partial r_2} \right |_{r_2=0} $ in the $(r_3,Y)$ plane. Evaluation directly in scale space (left) and from $\frac{1}{2}\left( {\frac{\partial U_1^+ }{\partial x_2^+}}+{\frac{\partial  U_1^- }{\partial x_2^-}}\right)$ (right)}
    \label{dudr2_com}
\end{figure}

Comparison with evaluation at points $x_i^+$ and $x_i^-$ for the standard-averaged streamwise velocity shows that the results are identical, refer to figure \ref{dudr2_com}.

\subsection{Calculation of $\left. \frac{\partial  dU_i^{(AA)}}{\partial r_1}\right|_{r_1=0}$}

\begin{figure}
    \centering
  \includegraphics[scale=0.4]{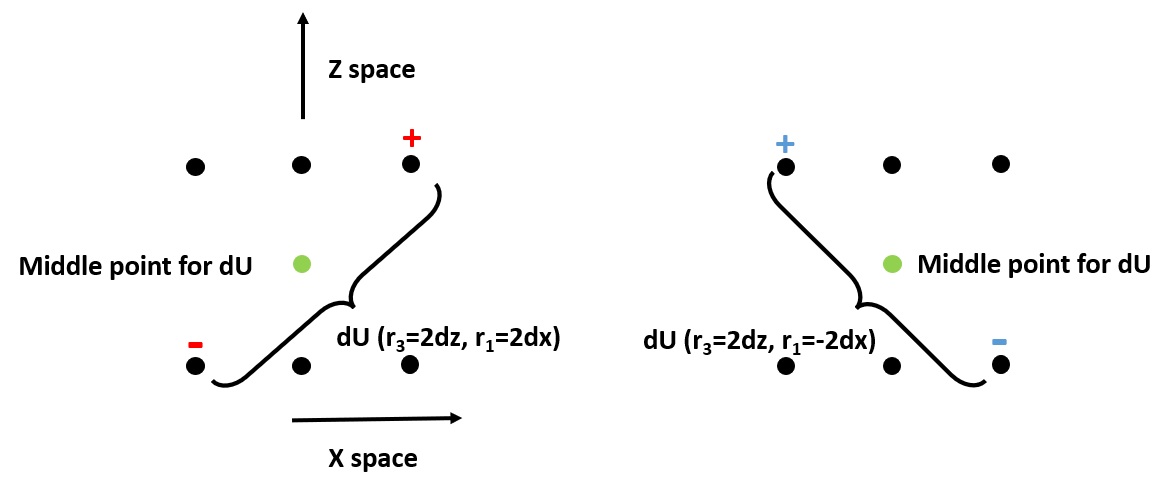}
    \caption{Sketch for the calculation of  $\left. \frac{\partial  dU_i^{(AA)}}{\partial r_1}\right|_{r_1=0}$.}
    \label{dudr1}
\end{figure}

\begin{figure}
    \centering
   \includegraphics[scale=0.4]{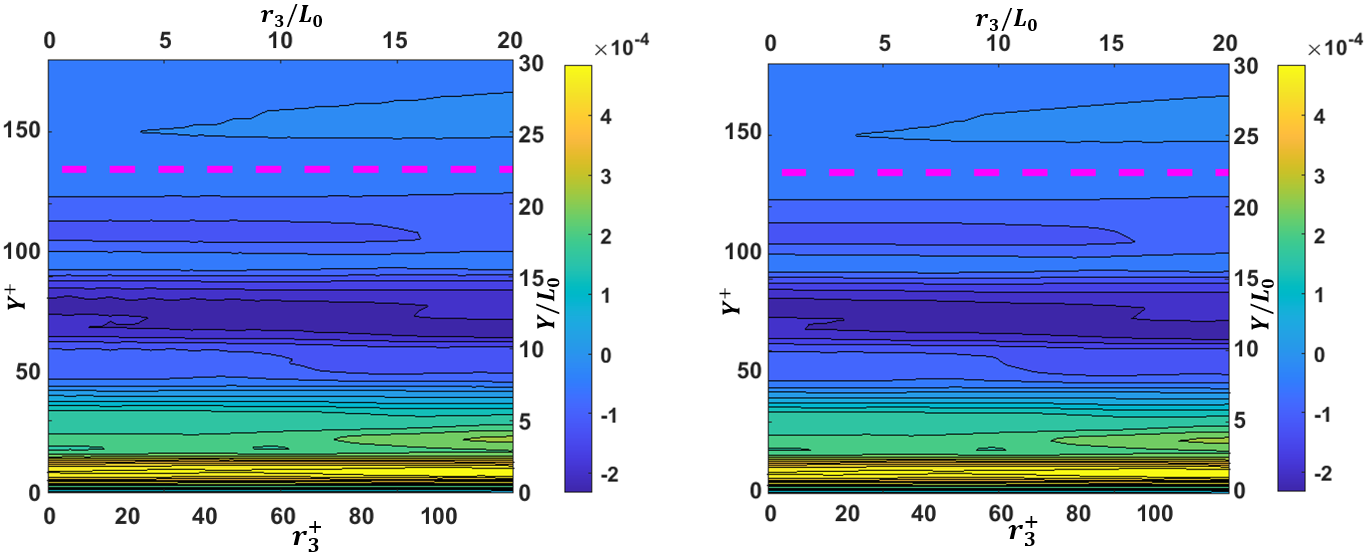}
    \caption{Contour plot of $\left. \frac{\partial  dU_1}{\partial r_1}\right |_{r_1=0} $ in the $(r_3,Y)$ plane. Evaluation directly in scale space (left) and from $\frac{1}{2}\left( {\frac{\partial U_1^+ }{\partial x_1^+}}+{\frac{\partial  U_1^- }{\partial x_1^-}}\right)$ (right).}
    \label{dudr1_com}
\end{figure}

\begin{enumerate}
\item Calculate $dU_i^{(AA)}(r_3,r_1=2\Delta x)$, i.e.\ the velocity difference between the red $+/-$ points shown in the left panel of figure \ref{dudr1}, at a fixed height $y=Y_0$. 
\item Calculate $dU_i^{(AA)}(r_3,r_1=-2 \Delta x)$,  i.e.\ the velocity difference between the blue $+/-$ points shown in right panel of figure \ref{dudr1}, at a fixed height $y=Y_0$.
\item Use central difference scheme to compute
\begin{equation}
\left. \frac{\partial  dU_i^{(AA)}}{\partial r_1}\right|_{r_1=0}=
\frac{ dU_i^{(AA)}(r_3,r_1=2 \Delta x) -dU_i^{(AA)}(r_3,r_1=-2\Delta x)}{2(2 \Delta x)}
\end{equation}
and store the value at the middle point (marked with a green dot in both panels of figure \ref{dudr1}).  
\item Repeat steps (i)-(iii) at different heights $Y_0$. 
\end{enumerate}
Schematic \ref{dudr1} shows the four points involved for $\left. \frac{\partial  dU_i^{(AA)}}{\partial r_1}\right|_{r_1=0}$. The process is similar for the evaluation of $ \frac{\partial  dU_i^{(AA)}}{\partial r_1}$ for different $r_1$ values.  Comparison with evaluation at points $x_i^+$ and $x_i^-$ for the standard-averaged streamwise velocity shows identical results, see figure \ref{dudr1_com}.

\section{Conditional decomposition of total fluxes in scale and physical spaces} \label{sec:appendix_total_fluxes}

\textbf{Decomposition of the total flux in scale space}
\begin{equation}
\boldsymbol{\phi_r}_j= \gamma^{(TT)}\boldsymbol{\phi_r}^{(TT)}_j+
\gamma^{(TL)}\boldsymbol{\phi_r}^{(TL)}_j+
\gamma^{(LT)}\boldsymbol{\phi_r}^{(LT)}_j+
\gamma^{(LL)}\boldsymbol{\phi_r}^{(LL)}_j+
\phi,
\label{}
\end{equation}
where
\begin{equation}
\boldsymbol{\phi_r}_j^{(AA)}=
\overline{ d u'^{(AA)}_j d q^{2^{(AA)}} }^{(AA)}
+
\overline{ dU_j^{(AA)} d q^{2^{(AA)}} }^{(AA)}
-
2\nu \overline{\frac{\partial d q^{2^{(AA)}} }{\partial r_j}}^{(AA)}
\label{}
\end{equation}

\begin{equation}
\begin{aligned}
\phi&=\gamma^{(TT)}\left( \overline{(du_i)^2}^{(TT)}dU_j^{(TT)}
+\overline{2du_idu_j}^{(TT)}dU_i^{(TT)}
-2dU_i^{(TT)}dU_i^{(TT)}dU_j^{(TT)} \right)\\
&
+
\gamma^{(TL)} \left(
\overline{(du_i)^2}^{(TL)}dU_j^{(TL)}
+\overline{2du_idu_j}^{(TL)}dU_i^{(TL)}
-2dU_i^{(TL)}dU_i^{(TL)}dU_j^{(TL)} \right) \\
&
+
\gamma^{(LT)}\left(
\overline{(du_i)^2}^{(LT)}dU_j^{(LT)}
+\overline{2du_idu_j}^{(LT)}dU_i^{(LT)}
-2dU_i^{(LT)}dU_i^{(LT)}dU_j^{(LT)} \right) \\
&
+
\gamma^{(LL)}\left(
\overline{(du_i)^2}^{(LL)}dU_j^{(LL)}
+\overline{2du_idu_j}^{(LL)}dU_i^{(LL)}
-2dU_i^{(LL)}dU_i^{(LL)}dU_j^{(LL)} \right)\\
&+
\gamma^{(TT)}dU_i^{(TT)}dU_i^{(TT)}dU_j
+
\gamma^{(TL)}dU_i^{(TL)}dU_i^{(TL)}dU_j\\
&
+
\gamma^{(LT)})dU_i^{(LT)}dU_i^{(LT)}dU_j
+
\gamma^{(LL)}dU_i^{(LL)}dU_i^{(LL)}dU_j
\\
&-2
\nu\gamma^{(TT)}\left(2dU_i^{(TT)}\frac{\partial  dU_i^{(TT)} }{\partial r_j}
+
2dU_i^{(TT)}\overline{ \frac{\partial du_i }{\partial r_j}}^{(TT)}
-
\frac{\partial  (dU_i^{(TT)})^2 }{\partial r_j}\right) \\
&
-2
\nu\gamma^{(TL)}\left(
2dU_i^{(TL)}\frac{\partial  dU_i^{(TL)} }{\partial r_j}
+
2dU_i^{(TL)}\overline{ \frac{\partial du_i }{\partial r_j}}^{(TL)}
-
\frac{\partial  (dU_i^{(TL)})^2 }{\partial r_j}\right) \\
&
-2
\nu\gamma^{(LT)}\left(
2dU_i^{(LT)}\frac{\partial  dU_i^{(LT)} }{\partial r_j}
+
2dU_i^{(LT)}\overline{ \frac{\partial du_i }{\partial r_j}}^{(LT)}
-
\frac{\partial  (dU_i^{(LT)})^2 }{\partial r_j}\right) \\
&
-2
\nu\gamma^{(LL)}\left(
2dU_i^{(LL)}\frac{\partial  dU_i^{(LL)} }{\partial r_j}
+
2dU_i^{(LL)}\overline{ \frac{\partial du_i }{\partial r_j}}^{(LL)}
-
\frac{\partial  (dU_i^{(LL)})^2 }{\partial r_j}\right)  \\
&-
\overline{(du_i)^2}dU_j
-\overline{2du_idu_j}dU_i
+dU_idU_idU_j 
-2
\nu\left(
-
2dU_i\frac{\partial  dU_i }{\partial r_j}
-
2dU_i\overline{ \frac{\partial du_i }{\partial r_j}}
+
\frac{\partial  (dU_i)^2 }{\partial r_j}
\right)
\label{eq:flux_scale_space_total_phi}
\end{aligned}
\end{equation}

\textbf{Decomposition of the total flux in physical space}
\begin{equation}
\boldsymbol{\phi_s}_j= \gamma^{(TT)}\boldsymbol{\phi_s}^{(TT)}_j +
\gamma^{(TL)}\boldsymbol{\phi_s}^{(TL)}_j+
\gamma^{(LT)}\boldsymbol{\phi_s}^{(LT)}_j+
\gamma^{(LL)}\boldsymbol{\phi_s}^{(LL)}_j+
\phi,
\label{}
\end{equation} 
where
\begin{equation}
\begin{aligned}
&\boldsymbol{\phi_s}_j^{(AA)}= \\
&\overline{{U^*_j}^{(AA)}d q^{2^{(AA)}}}^{(AA)}
+
\overline{{u'^*_j}^{(AA)}d q^{2^{(AA)}}}^{(AA)} 
+
\overline{2 d u'^{(AA)}_j d p'^{(AA)} }^{(AA)} 
-
\frac{1}{2} \nu  \overline{\frac{\partial  d q^{2^{(AA)}} }{\partial X_j}}^{(AA)}
\label{}
\end{aligned}
\end{equation}

\begin{equation}
\begin{aligned}
\phi&=\gamma^{(TT)}\left( \overline{(du_i)^2}^{(TT)}{U^*_j}^{(TT)}
+\overline{2du_i{U^*_j}}^{(TT)}dU_i^{(TT)}
-2dU_i^{(TT)}dU_i^{(TT)}{U^*_j}^{(TT)} \right)\\
&
+
\gamma^{(TL)} \left(
\overline{(du_i)^2}^{(TL)}{U^*_j}^{(TL)}
+\overline{2du_i{U^*_j}}^{(TL)}dU_i^{(TL)}
-2dU_i^{(TL)}dU_i^{(TL)}{U^*_j}^{(TL)} \right) \\
&
+
\gamma^{(LT)}\left(
\overline{(du_i)^2}^{(LT)}{U^*_j}^{(LT)}
+\overline{2du_i{U^*_j}}^{(LT)}dU_i^{(LT)}
-2dU_i^{(LT)}dU_i^{(LT)}{U^*_j}^{(LT)} \right) \\
&
+
\gamma^{(LL)}\left(
\overline{(du_i)^2}^{(LL)}{U^*_j}^{(LL)}
+\overline{2du_i{U^*_j}}^{(LL)}dU_i^{(LL)}
-2dU_i^{(LL)}dU_i^{(LL)}{U^*_j}^{(LL)} \right)\\
&+
\gamma^{(TT)}dU_i^{(TT)}dU_i^{(TT)}U^*_j
+
\gamma^{(TL)}dU_i^{(TL)}dU_i^{(TL)}U^*_j\\
&
+
\gamma^{(LT)}dU_i^{(LT)}dU_i^{(LT)}U^*_j
+
\gamma^{(LL)}dU_i^{(LL)}dU_i^{(LL)}U^*_j
\\
&-\frac{1}{2}
\nu\gamma^{(TT)}\left(2dU_i^{(TT)}\frac{\partial  dU_i^{(TT)} }{\partial X_j}
+
2dU_i^{(TT)}\overline{ \frac{\partial du_i }{\partial X_j}}^{(TT)}
-
\frac{\partial  (dU_i^{(TT)})^2 }{\partial X_j}\right) \\
&
-\frac{1}{2}
\nu\gamma^{(TL)}\left(
2dU_i^{(TL)}\frac{\partial  dU_i^{(TL)} }{\partial X_j}
+
2dU_i^{(TL)}\overline{ \frac{\partial du_i }{\partial X_j}}^{(TL)}
-
\frac{\partial  (dU_i^{(TL)})^2 }{\partial X_j}\right) \\
&
-\frac{1}{2}
\nu\gamma^{(LT)}\left(
2dU_i^{(LT)}\frac{\partial  dU_i^{(LT)} }{\partial X_j}
+
2dU_i^{(LT)}\overline{ \frac{\partial du_i }{\partial X_j}}^{(LT)}
-
\frac{\partial  (dU_i^{(LT)})^2 }{\partial X_j}\right) \\
&
-\frac{1}{2}
\nu\gamma^{(LL)}\left(
2dU_i^{(LL)}\frac{\partial  dU_i^{(LL)} }{\partial X_j}
+
2dU_i^{(LL)}\overline{ \frac{\partial du_i }{\partial X_j}}^{(LL)}
-
\frac{\partial  (dU_i^{(LL)})^2 }{\partial X_j}\right)  \\
&+2\left(
\gamma^{(TT)}dU_i^{(TT)}dP^{(TT)}+
\gamma^{(TL)}dU_i^{(TL)}dP^{(TL)}+
\gamma^{(LT)}dU_i^{(LT)}dP^{(LT)}+ \right.\\
&\qquad
\left. \gamma^{(LL)}dU_i^{(LL)}dP^{(LL)}\right)
-\frac{1}{2}
\nu\left(
-
2dU_i\frac{\partial  dU_i }{\partial X_j}
-
2dU_i\overline{ \frac{\partial du_i }{\partial X_j}}
+
\frac{\partial  (dU_i)^2 }{\partial X_j}
\right)\\
&-
\overline{(du_i)^2}U^*_j
-\overline{2du_iU^*_j}dU_i
+dU_idU_iU^*_j-2dU_idP\\
\label{eq:flux_physical_space_total_phi}
\end{aligned}
\end{equation}

\bibliographystyle{jfm}
\bibliography{jfm-instructions,References_GP}

\end{document}